\documentstyle[pre,aps,eqsecnum,amssymb,multicol,twoside,rotate,epsf]{revtex}

\begin{document}

\draft

\title{\Large\bf Phase diagrams, critical and multicritical behavior \\
  of hard-core Bose-Hubbard models}

\author{Christian Pich$^{a}$ and Erwin Frey$^{b}$}

\address{$^{a}$Physics Department, University of California, Santa Cruz, CA
  95064 \\ $^{b}$Institut f\"ur Theoretische Physik, Technische
  Universit\"at M\"unchen, D-85747 Garching, Germany}

\date{\today}

\maketitle

\begin{abstract}
  We determine the zero-temperature phase diagram of the hard-core Bose-Hubbard
  model on a square lattice by mean-field theory supplemented by a linear
  spin-wave analysis. Due to the interplay between nearest and next-nearest
  neighbor interaction and cubic anisotropy several supersolid phases with
  checkerboard, stripe domain or intermediate symmetry are stabilized. The
  phase diagrams show three different topologies depending on the relative
  strength of nearest and next-nearest neighbor interaction. We also find a
  rich variety of new quantum critical behavior and multicritical points and
  discuss the corresponding effective actions and universality classes.
\end{abstract}


\begin{multicols}{2}
\narrowtext

\section{Introduction}
\label{section:introduction}

Since the original suggestion by Andreev and Lifshitz~\cite{andreev:69},
Chester~\cite{chester:70} and Leggett~\cite{leggett:70} that quantum crystals
such as $^4$He might exhibit a phase where superfluidity coexists with
crystalline order the topic of supersolids has received considerable
theoretical attention. Early theoretical work (for a review see e.g.
Ref.~\cite{andreev-review:82}) focused on the possible implications of large
zero-point vibrations in establishing a supersolid phase in quantum crystals.
It was argued by Andreev and Lifshitz~\cite{andreev:69} that these quantum
effects might be sufficient to delocalize either impurities and/or zero-point
vacancies. Such a system would form a weakly interacting Bose gas and therefore
be a beautiful example of a system which shows Bose-Einstein condensation. Upon
exploiting a mapping between hard-core lattice gas models and spin-$1/2$
Heisenberg models it was shown that within a mean-field approximation such a
supersolid phase exists for systems with a finite range of interactions between
the bosons~\cite{matsuda:70,liu:73}.

Unfortunately, there is not yet clear experimental evidence for such a phase.
The most promising candidates for experimental systems, where supersolid order
might be observed, are Josephson junction arrays and $^4$He films on substrates
in two dimensions. There have also been experimental searches for supersolid
order in bulk $^4$He. In ultrasound studies~\cite{lengua-goodkind:90} of highly
purified solid $^4$He a recently observed resonance phenomenon was interpreted
to be consistent with the presence of a supersolid induced by zero-point
vacancies. But the experimental situation still remains
controversial~\cite{meisel:92}, and additional evidence is needed to
unambiguously prove the existence of such a phase. Artificially fabricated
Josephson junction arrays are a particular interesting system for the
observation of exotic phases in quantum systems.  Finite-range interactions and
frustrations present in these arrays give rise to a rather rich structure of
the phase diagram.

Partly motivated by this controversial experimental situation, recent work has
concentrated on mapping out the phase diagram of strong interacting clean
bosonic systems mainly through numerical and mean-field approximation
\cite{bruder:93,roddick:93,otterlo:94,scalettar:95,batrouni:95,otterlo:95}. The
starting point of these investigations are lattice models of interacting bosons
with the following Bose-Hubbard Hamiltonian,
\begin{eqnarray}
  {\cal H} = -\frac{t}{2} \sum_{\langle i,j \rangle} 
      \left(a_i^\dagger a_j  + a_j^\dagger a_i \right) 
      - \mu \sum_i n_i 
 + \sum_{i,j} n_i U_{ij} n_j .
\label{bose-hubbard-hamiltonian}
\end{eqnarray}
Here $a_i$, $a_i^\dagger$ are boson annihilation and creation operators at site
$i$, and $n_i = a_i^\dagger a_i$. The hopping integral $t$ sets the energy
scale, $\mu$ is the chemical potential, and $U_{ij}$ denotes the interaction
between the bosons. Bosons only hop to adjacent places ($\langle i,j\rangle$).
It is found that for nearest neighbor interaction supersolid order at
half-filling is favored at large nearest neighbor interaction $U_1$ and small
on-site interaction $U_{ii} = U_0$, but becomes suppressed in the hard-core
limit~\cite{otterlo:95}. In the latter case next-nearest neighbor interactions
become necessary for supersolid order to exist~\cite{liu:73}. Besides the
quantum ground states at half filling, which may be either a checkerboard or
stripe domain density wave, additional Mott-insulating phases with a density
wave commensurable with the underlying lattice appear.  Each of these phases
seems to have an associated supersolid phase in which the particular
charge-density order coexists with off-diagonal long range order. As we will
show here, however, there are intermediate supersolid phases which have no
Mott-insulating partner in the phase diagram.

In the present work we report on a mean-field analysis of the hard-core
next-nearest neighbor Bose-Hubbard model supplemented by a linear spin-wave
analysis. Our purpose is to map out the phase diagram over the whole range of
parameters of nearest and next-nearest interaction. This extends previous work
by Bruder et al.~\cite{bruder:93}, which has been restricted to a small value
of the next-nearest neighbor interaction, where the ground state possesses
checkerboard symmetry. Furthermore, we determine the order and universality
class of the phase transitions and identify various interesting multicritical
points.

We focus on strong on-site interaction $U_{ii} = U_0$, i.e.\ a hard-core
approximation where the particle number per site takes only values between the
integer values $n$ and $n+1$ (in the vicinity of half-filling). Then the model
can be transformed in a spin-$1/2$ XXZ model \cite{matsuda:70}
\begin{eqnarray}
{\cal H} = &&-J\sum_{\langle i,j \rangle}\left(S_i^xS_j^x+S_i^yS_j^y\right)
+U_1\sum_{\langle i,j \rangle}S_i^zS_j^z \nonumber \\
&& +U_2\sum_{\langle\langle i,j \rangle\rangle} S_i^zS_j^z - h\sum_i S_i^z\, ,
\label{XXZ}
\end{eqnarray}
where we have dropped a constant energy offset.  The mapping is performed by
identifying $a_i^\dagger = S_i^x + i S_i^y$, $n_i = S_i^z + \frac{1}{2}$, $J =
t$ and $h = \mu - \sum_i U_{0i}$. Here $\langle i,j\rangle$ denotes the nearest
neighbors and $\langle\langle i,j\rangle\rangle$ the next-nearest neighbors.
After a Fourier transform, $S_i^\alpha = \frac{1}{\sqrt{N}} \sum_q e^{i q x_i}
S_q^\alpha$, the Hamiltonian can be written as 
\begin{eqnarray}
{\cal H} = -\sum_q \Biggl[\!&&J_q \left(S_q^xS_{-q}^x+S_q^yS_{-q}^y\right) 
\nonumber \\
&&+(U_{1q}+U_{2q})S_q^zS_{-q}^z \Biggr]-\sqrt N hS_{q=0}^z
\end{eqnarray}
with the interaction terms
\begin{mathletters}
\begin{eqnarray}
J_q & = & z_1J\gamma_q, \\
U_{1q} & = & -z_1U_1 \gamma_q, \\
U_{2q} & = & -z_2U_2 \eta_q.
\end{eqnarray}
\end{mathletters}
The structure factors 
\begin{eqnarray}
\gamma_q = \frac{1}{z_1} \sum_{l\in\text{n.n.}} e^{iqx_l}, 
\quad \text{and} \quad
\eta_q  = \frac{1}{z_2} \sum_{l\in \text{n.n.n.}} e^{iqx_l}
\end{eqnarray}
depend on the lattice structure and are defined as sums over the nearest
neighbor sites $z_1$ and next-nearest neighbor sites $z_2$, respectively.

In this paper we use classical ground state analysis and linear spin-wave
theory to explore the phase diagram of the hard-core Bose-Hubbard system with
nearest and next-nearest neighbor ``charge''-interaction, $U_{ij}$, and a
nearest neighbor hopping term, $J$. We focus our study on the bipartite square
lattice. For a discussion of frustrated two-dimensional lattices, like the
triangular and Kagom\'e lattice, we refer the reader to a recent paper by
Murthy et al.~\cite{murthy-arovas-auerbach:97}.

Linear spin-wave theory is implemented in its standard form. First, one
calculates the classical ground state energy and the corresponding spin
configuration, which frequently may be described by some ordering wave vector
$\tilde q$. Next, local rotations of the spins are performed, $S_i^\alpha
\rightarrow {\tilde S_i^\alpha}$, such that the $z$-component $\tilde S_i^z$
points along the direction of the classical spin configuration. Upon performing
a Holstein-Primakoff transformation, which in Fourier space is given by (for
those ground states which can be described by a single wave vector $\tilde q$)
\begin{mathletters}
\begin{eqnarray}
\tilde S_q^x = \sqrt\frac{S}{2}(a_{-q}+a^\dagger_q), \\
\tilde S_q^y = -i\sqrt\frac{S}{2}(a_{-q-\tilde q}-a^\dagger_{q+\tilde q}), \\
\tilde S_q^z = \sqrt N S\delta_{q,\tilde q} -\frac{1}{\sqrt N}
               \sum_p  a^\dagger_{p-q-\tilde q} a_p, 
\label{Trafo}
\end{eqnarray}
\end{mathletters}
a linearized spin-wave Hamiltonian ${\cal H}_{\text{sw}}$, describing the
elementary excitation around the classical ground state, is obtained.

The outline of the paper is as follows. In section II we study the phase
diagram of the hard-core Bose-Hubbard model with nearest and next-nearest
neighbor interaction. The analysis is done by employing the mapping to the
spin-$1/2$ XXZ model and using a mean-field analysis supplemented by a linear
spin-wave analysis. This allows us to determine the ground state spin
configurations in the spin-$1/2$ XXZ model and analyze its stability and
soft-modes by linear spin-wave theory. The resulting phase diagrams are given
in section III together with a discussion of the universality classes of the
critical phenomena at the various phase boundaries.  We also identify a series
of multicritical points and discuss their multicritical behavior. Our results
and conclusions are given in section IV.  Finally, some technical details about
the linear spin-wave analysis of the Mott insulating 3/4 phase and the
checkerboard and stripe domain supersolid phases are deferred to the
appendices.

\section{Mean-field and linear spin-wave theory}
\label{section:mean_field}

\begin{figure}[hbt]
  \epsfxsize = 0.75\columnwidth \centerline{
    \epsffile{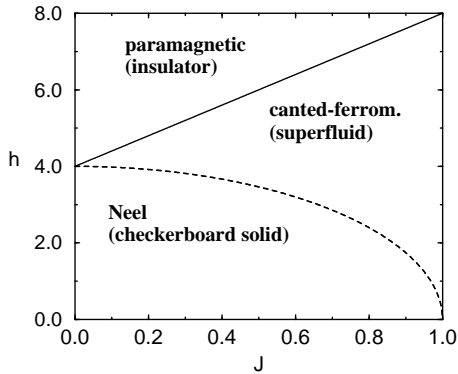} }
\vspace{0.1cm}
\caption{Mean-field phase diagram of a two-dimensional hard-core Bose-Hubbard
  with nearest neighbor interaction $U_1=1$ and $S=1/2$. The N\'eel,
  canted-ferromagnetic and 
  paramagnetic phase in the spin-$1/2$ XXZ model correspond to a Mott
  insulating phase with a checkerboard charge ordering, superfluid and a
  insulating phase with a homogeneous charge density, respectively. Second and
  first order phase transitions are indicated by solid and dashed lines,
  respectively.}
\label{fig:pd_nearest}
\end{figure}

In this section we analyze the phase diagram of the next-nearest neighbor
hard-core Bose-Hubbard model, using a mean-field analysis supplemented by a
linear spin-wave analysis. However, before discussing the effect of
next-nearest neighbor interactions let us shortly summarize the topology of the
phase diagram when the interaction $U_{ij}$ is restricted to nearest neighbors.
The phase diagram of the spin-$1/2$ XXZ Heisenberg model becomes particularly
simple~\cite{bruder:93} (s.\ Fig.~\ref{fig:pd_nearest}).  There is a
``half-integer'' lobe (N\'eel phase) centered around zero magnetic field
($J<U_1$). In the original Bose-Hubbard model this corresponds to the
half-filling Mott-insulating phase with a checkerboard charge density wave. For
increasing magnetic field (away from half-filling) this phase becomes unstable
to a canted ferromagnetic state, i.e. the corresponding bosonic system becomes
superfluid.  Finally, for strong fields the system goes into a paramagnetic
state with magnetization pointing in the field direction. Such a state has
uniform density of bosons and is insulating again.

Taking into account longer-range interactions, the phase diagram acquires
further structure. In addition to the phases discussed above, one finds more
general insulating phases and superconducting phases in which the spins arrange
in a $2\times2$ unit cell. In particular one can identify regions in parameter
space where coexistence of long-range order in the $xy$-direction and staggered
magnetization in the $z$-direction appear. This implies that the corresponding
bosonic system would be a supersolid, since there is coexistence of crystalline
order and superfluidity. A recent analysis of the classical ground state
energies by Bruder et al.~\cite{bruder:93} identifies two different supersolid
phases, which are characterized by two or three angles in the $2\times2$ unit
cell. However, their analysis has been restricted to small values of the
next-nearest neighbor interaction, where the ground state possesses
checkerboard symmetry. Here we extend the mean-field analysis of
Ref.~\cite{bruder:93} to the full range of the parameters, where the ground
state of the Bose-Hubbard model at half-filling possesses stripe symmetry.  In
addition we supplement the analysis of the classical ground state energy by a
linear spin-wave analysis. In performing a spin-wave analysis we can check
whether the ground state found by energy considerations is stable. In general,
one has to minimize within a small symmetry class (an Ansatz), i.e. we make an
Ansatz for the number of sublattices we need.  From stability investigations we
can confirm the ground state analysis, and we can evaluate the boundaries to
the adjacent phases which occur as a soft-mode. In addition, the wave vector of
the soft-mode gives a hint of the adjacent phase structure. Moreover, we obtain
the contribution of the quantum fluctuations to the reduction of the true
ground state energy and the order parameter.

Now we turn to a linear spin-wave analysis of the model, taking into account
next-nearest neighbor interaction. For that we need the wave vector dependence
of the interaction parameters in the linear spin-wave Hamiltonian.  On a square
lattice ($xy$-plane) there are four nearest and four next-nearest neighbors
($z_1 = z_2 = 4$) and the structure factors are given by ($a=1$)
\begin{eqnarray}
\gamma_q & = & \frac{1}{2}(\cos q_x +\cos q_y) , \\
\eta_q & = & \frac{1}{2}(\cos{(q_x+q_y)} +\cos{(q_x-q_y)} ) 
        = \cos q_x\cos q_y . 
\end{eqnarray}
We proceed in section~\ref{collinear_phases} by a soft-mode analysis of the
collinear phase found in the nearest neighbor model. This allows us to identify
the stability boundaries of these ground states. Together with a soft-mode
analysis of the other Mott-insulating phases with quarter-filling in section
\ref{quarter_mott_insulating}, this gives the regions in parameter space where
intermediate phases (supersolid phases) are to be expected. Using an Ansatz
with a general spin configuration on a $2\times2$ unit cell leads us in section
\ref{non_collinear} to an identification of the symmetry of the order
parameters in the corresponding non-collinear phases.

\subsection{Collinear phases}
\label{collinear_phases}

In this section we investigate the collinear phases, i.e. those phases in which
the spins are parallel. These are the antiferromagnetic phases, namely the
N\'eel and stripe phase, the paramagnetic phase and the canted-ferromagnetic
phase.

\subsubsection{Antiferromagnetic phases: \\ 
  N\'eel and stripe domain ground states}
\label{neel_stripe}

For vanishing fields, i.e. half-filling, we have a N\'eel, stripe or a
ferromagnetic phase depending on the relative strengths of the interaction
parameters $U_1, U_2$ and $J$.  In the antiferromagnetic phases (N\'eel or
stripe) the spins are oriented along the $z$-axis, which corresponds
to an Ising-like anisotropy.  Anticipating such a commensurate structure with
ordering wave vector $\tilde q $, which is $q_0 = \pi (1,1)$ for the N\'eel
and $q_1 = \pi (0,1)$ or $q_2 = \pi (1,0)$ for the stripe phase, a
Holstein-Primakoff transform yields the spin-wave Hamiltonian
\begin{eqnarray}
{\cal H}_{\text{sw}} = E_g 
+\sum_q \Biggl[\!&&A_q a^\dagger_qa_q\ +\frac{B_q}{2}(a_qa_{-q}+a^\dagger_q
a^\dagger_{-q}) \nonumber \\
&&+\frac{h}{2}(a_q^\dagger a_{q+\tilde q}+a_{q+\tilde q}^\dagger a_q) 
\Biggr]. 
\label{sw_square_lattice}
\end{eqnarray}
The coefficients of the Hamiltonian read
\begin{mathletters}
\begin{eqnarray}
  A_q & = & S[2(U_{1\tilde q}+U_{2\tilde q})-J_q-J_{q+\tilde q}] ,\\
  B_q & = & S(J_{q+\tilde q}-J_q)\, .
\end{eqnarray}
\end{mathletters}
For the N\'eel phase there are only two ground states, whereas four ground
states with the same classical energy exist for the stripe phase.
Diagonalization of the spin-wave Hamiltonian, Eq.~\ref{sw_square_lattice},
gives the classical ground state energies
\begin{eqnarray}
  E_g = \cases { -4NS^2(U_1-U_2), & N\'eel \cr
                 -4NS^2U_2, & Stripe\cr }\, ,
\label{gs_energy_af}
\end{eqnarray}
and the corresponding two branches of the spin-wave spectra $\epsilon_q^{\pm}
= \epsilon_q \pm h$ with 
\begin{equation}
\epsilon_q= 8 S \cases {\sqrt{(U_1-U_2)^2-J^2\gamma_q^2} & 
                       \cr
                       \sqrt{(U_2-\frac{J}{2}\cos q_x)^2-\frac{1}{4}J^2
                       \cos^2\! q_y} &\cr}\, ,
\label{sws_square_lattice}
\end{equation}
where we have picked the ordering wave vector $q_1$ for the stripe domain
phase. If one takes $q_2$ instead, then the corresponding excitation spectrum
is obtained from Eq.~\ref{sws_square_lattice} by interchanging $q_x$ with
$q_y$.

Thus, when comparing the ground state energies, there is a transition from the
N\'eel state to the stripe state at
\begin{equation}
U_2 = U_1 / 2  .
\end{equation}
Since the two ground states have different symmetry the transition is
discontinuous (there is no intermediate spin orientation in a mean field
approximation and from spin-wave analysis). The domain of stability of each of
these classical ground states can be estimated from a soft-mode analysis within
linear spin-wave theory. The energy needed to excite a spin-wave at wave vector
$q$ in the collinear phases is given by Eq.~\ref{sws_square_lattice}. By
increasing the field $h$ a point can be reached where the lower branch of the
excitation energy vanishes at a certain wave vector $q$.  As a consequence, the
spin-wave at this particular wave vector becomes soft, the classical ground
state under consideration becomes unstable, and a phase transition to a new
phase takes place. Here the minimum in the lower branch of the excitation
energies is reached at the Brillouin zone center ($q=0$) for both ground
states. The corresponding upper stability boundaries for the antiferromagnetic
(af) zero-field ground states are given by
\begin{equation}
h_{\text{af}}^{\text{max}} = \cases{ 8S\sqrt{(U_1-U_2)^2-J^2}, & N\'eel \cr
               8S\sqrt{U_2(U_2-J)}, & Stripe\cr }.
\label{af_critical_field}
\end{equation}
For fields $h$ lower than $h^{\text{max}}_{\text{af}}$ the N\'eel and stripe
domain phases are stable against quantum fluctuations. Note that in the stripe
phase the parameter $U_1$ does not occur explicitly in the ground state energy
and in the spin-wave spectrum. 

For $h<h_{\text{af}}^{\text{max}}$ the dispersion relation shows an energy gap
for both phases because there is no continuous degeneracy of the ground state.
Thus, these phases have long-range order even at finite temperatures. At $T=0$
the order parameter (staggered magnetization) is reduced by quantum
fluctuations
\begin{equation} 
{\rm N} = g\mu_B NS \left[1-\frac{1}{2NS}\sum_q 
    \left(\frac{A_q}{\epsilon_q}-1\right)\right]\, ,
\end{equation}
with $\epsilon_q$ from Eq. \ref{sws_square_lattice}. From the latter we
conclude that this reduction is independent from the field, too. Calculating
the magnetization we readily see that there is no contribution from quantum
fluctuations; thus, the magnetization vanishes in the whole lobe. This
corresponds to a constant mean particle number $\langle n\rangle \simeq \langle
S^z\rangle$, which is a signature of a Mott insulator. The tip of the lobes of
the antiferromagnetic phases can be found from the zeros of the critical
fields, Eq.~\ref{af_critical_field}. The corresponding limits for the hopping
integrals are
\begin{equation}
J < \cases { U_1-U_2, & \text{for checkerboard symmetry,} \cr
             U_2,     & \text{for stripe symmetry}.\cr }
\label{H_c-afm}
\end{equation}

\subsubsection{Paramagnetic phase}

In the paramagnetic phase all spins are oriented along the field direction
($z$-direction). Applying the Holstein-Primakoff transform we get the following
spin-wave Hamiltonian
\begin{equation}
{\cal H}_{\text{sw}} = E_g +\sum_q \epsilon_q a^\dagger_qa_q\, ,
\end{equation}
with the classical ground state energy $E_g$ and the spin-wave spectrum
$\epsilon_q$
\begin{eqnarray}
E_g & = & 4 NS^2(U_1+U_2) -NSh, \\
\epsilon_q & = & h - 2S(U_{10}+U_{20}+J_q).
\label{gs_energy_para}
\end{eqnarray}
Stability of the ground state requires a positive excitation spectrum which
gives the lower bound of the paramagnetic phase (soft-mode for $q=0$)
\begin{equation}
h_{\text{min}}^{\text{para}} = h_c = 8S(U_1+U_2+J).
\label{para_critical_field}
\end{equation}

\subsubsection{Canted-ferromagnetic phase}

For vanishing field and large values of $J$ the spins align ferromagnetically
in the plane. Due to the rotation symmetry in the $xy$-plane there is a
Goldstone mode.  For an infinitesimal field perpendicular to the plane this
ferromagnetic phase becomes unstable and changes to a canted phase where the
spins orient ferromagnetically towards the field direction. By minimizing the
energy we get a relation between the canting angle and the magnetic field
\begin{equation}
h = h_c\sin\theta \, .
\end{equation}
Here $\theta$ denotes the angle between the plane and the spin direction. The
resulting ground state energy can be written as
\begin{equation}
E_g = -4NS^2J-NS{h^2\over 2h_c} \, ,
\label{gs_energy_canted}
\end{equation}    
and the dispersion relation for the canted-ferromagnetic state is
\begin{eqnarray}
\epsilon_q^2 = &4&S^2(J_0-J_q) 
               \Bigl[J_0-J_q \nonumber \\
               &&+(J_q+U_{1q}+U_{2q})\left(1-\frac{h^2}{h_c^2} \right)\Bigr].
\end{eqnarray}
By comparing the classical ground state energy of the paramagnetic and the
canted-ferromagnetic phase one finds that the transition to the paramagnetic
phase takes place at $h=h_c$. This agrees with the lower stability boundary,
Eq.~\ref{para_critical_field}, obtained from a soft-mode analysis of the
paramagnetic phase. At the transition the canting angle $\theta$ continuously
goes to $\pi/2$. Thus, in a mean-field approximation the corresponding phase
transition at $h=h_c$ is of second order and belongs to the $D=3$ classical XY
universality class.

The lower field boundary can again be obtained from a soft-mode analysis of the
excitation spectrum. One finds that depending on the relative magnitude of the
next-nearest and nearest neighbor interaction there is a soft-mode either at
the ordering wave-vector of the N\'eel state $q=q_0$ or of the stripe domain
state $q=q_1$, $q_2$. This gives
\begin{mathletters}
\begin{eqnarray}
h_{\text{min}}^{\text{N\'eel}} 
& = & h_c\sqrt\frac{{U_1-U_2-J}}{{U_1-U_2+J}}, \quad {\rm for}\quad  q_0, \\
\label{h_c2}
& \simeq & 8S(U_1+U_2) -16S{U_2\over U_1-U_2}J \, , 
\label{h_c2exp} 
\end{eqnarray}
\end{mathletters}
and 
\begin{mathletters}
\begin{eqnarray}
h^{\text{stripe}}_{\text{min}} & = 
& h_c\sqrt\frac{U_2-J}{U_2}, \quad {\rm for} \quad q_1, \, q_2.\\
\label{h_c21}
& \simeq & 8S(U_1+U_2)+4S{U_2-U_1\over U_2}J\, .
\label{h_c21exp}
\end{eqnarray}
\end{mathletters}
At these boundaries the canted-ferromagnetic phase becomes unstable to an
intermediate phase with a checkerboard or stripe-like spin configuration which
will be discussed in section~\ref{non_collinear}. Here we added the expansion
for small $J$.

\subsection{The Mott-insulating 3/4 and 1/4 lobe}
\label{quarter_mott_insulating}

In the preceding section we have seen that there must be intermediate phases
with a more complicated spin configuration. As shown in previous work
\cite{bruder:93}, there exist phases with non-integer filling, namely 3/4 and
1/4, i.e. 3 up and 1 down spin and vice versa for a reversed magnetic field.
The spins have no component in the plane. Such a ground state on a square
lattice corresponds to a {\em four-sublattice} system for which the ground
state is uniquely defined. By symmetry arguments the down spin (3/4 lobe) can
be at each position without changing the energy. The classical ground state of
this configuration is given by
\begin{equation}
E_g = -\frac{1}{2}NSh\, ,
\label{gs_energy_3/4}
\end{equation}
which is independent of the exchange energy $J$, $U_1$ and $U_2$. Performing a
spin-wave analysis we can calculate the excitation energy of four branches.
From stability conditions, i.e. positiveness of the excitation energies, we can
deduce various equations. For $J=0$ the Hamiltonian is already diagonal and the
four branches of the dispersion relation are given by (s. appendix)
\begin{mathletters}
\begin{eqnarray}
\epsilon_q^{(1)} & =  & h-8SU_2, \\
\epsilon_q^{(2)} & =  & h+8S(U_2-U_1), \\
\epsilon_q^{(3)} & =  & -h+8S(U_2+U_1),\\
\epsilon_q^{(4)} & =  & h-8SU_2.
\end{eqnarray}
\end{mathletters}
These excitations are independent of the wave vector $q$. From these equations
we get an upper and two lower bounds of the 3/4-lobe, 
depending on the relative strength of $U_1$ and $U_2$:
\begin{equation}
  8S(U_1+U_2) \geq h \geq \cases{ 8S(U_1-U_2), & $U_1 > 2U_2$\cr 
                                  8SU_2, & $U_1 <2U_2$}\, .
\end{equation}
The 3/4 phase vanishes for vanishing $U_2$. The location of the lower bounds
are different, depending on whether the zero-field ground state possesses
checkerboard or stripe domain symmetry. For finite $J$ one has to solve the
full dynamical matrix. Vanishing of this spectrum at $q=0$ leads to the
following condition (see appendix~\ref{app:sw_3/4})
\begin{eqnarray}
&&h^3-8S(2U_1+U_2)h^2+(8S)^2(U_1^2-U_2^2+2U_1U_2)h \nonumber \\
&&+(8S)^3U_2(U_2^2-U_1^2+J^2) = 0,
\label{sw_3/4}
\end{eqnarray}
which determines three solutions.  The fourth solution defines the boundary
$h=8SU_2$, which corresponds to $\epsilon^{(4)}$ and is independent of
$J$. To see which solutions are relevant we expand them for small $J$ values:
\begin{mathletters}
\begin{eqnarray}
h^{(1)} & \simeq & 8SU_2 +{8SU_2\over U_1(2U_2-U_1)}J^2\\
h^{(2)} & \simeq & 8S(U_2-U_1) +{4S\over U_1-2U_2}J^2\\
h^{(3)} & \simeq & 8S(U_1+U_2) -{4S\over U_1}J^2\\
h^{(4)} & = & 8SU_2\, .
\end{eqnarray}
\end{mathletters}
From these expansions we see that $h^{(3)}$ defines the upper boundary and
$h^{(1)}$ and $h^{(2)}$ the lower boundaries for the stripe and the N\'eel
ground states respectively. $h^{(4)}$ is irrelevant.  As will be shown in the
next section, the lower boundaries $h^{(1)}$ and $h^{(2)}$ define a continuous
transition to a non-collinear three-sublattice structure (SS2).

In comparison with the phase boundaries for the collinear spin phases
(Eq.~\ref{H_c-afm}) we readily see that the antiferromagnetic phases only meet
the $3/4$ phase at $J=0$.  However, the upper boundary needs some more
investigation: the phase boundaries from the canted-ferromagnetic to a
non-collinear phase (Eqs.~\ref{h_c2} and \ref{h_c21}) show that the $3/4$ phase
intrudes (for $U_2<U_1$) in the canted ferromagnetic phase. The solution of
this seemingly paradoxical situation is that there is a discontinuous
transition from $3/4$ to the canted-ferromagnetic phase, (for low $J$) which is
given by the equality of their ground state energies:
\begin{eqnarray}
  h_{\text{c3}} &=& 4S (U_1 + U_2 + J) \nonumber \\
  &&+ 4 S \sqrt{(U_1+U_2+J)(U_1+U_2-3J)}\\
\label{h_c3}
&\simeq & 8(U_1+U_2) -{8S\over U_1+U_2}J^2\, .
\label{cant-34} \nonumber
\end{eqnarray}
Thus, for $U_2<U_1$ this discontinuous transition is prefered and the upper
boundary ($h^{(3)}$) defines a continuous transition to a 
supersolid phase (SS2) only for
$U_2>U_1$.

\subsection{Non-collinear phases: supersolids}
\label{non_collinear}

\begin{figure}[htb]
  \epsfxsize=0.65\columnwidth
  \centerline{\epsffile{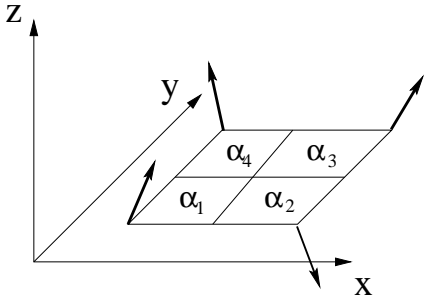}}
  \vspace{0.3cm}
\caption{General spin configuration with four different angles $\alpha_i$,
  $i=1, \cdots, 4$, in a $2\times2$ unit cell.}
\label{fig:2by2}
\end{figure}

The stability analysis in the preceding sections has shown that there must be
several non-collinear phases between the paramagnetic and the N\'eel/stripe
phase in which the spins have both a ferromagnetic and an antiferromagnetic
orientation. In order to describe these intermediate phases we use an Ansatz
with a general spin configuration with four different angles $\alpha_i$ between
the $z$-axis and the $xy$-plane, $i=1, \cdots, 4$, in a $2\times2$ unit cell
(see Fig.~\ref{fig:2by2}).
Due to the planar symmetry only these angles are independent.
In such a general four-sublattice model the ground state energy is given by
\begin{eqnarray}
  E_g &=& -{NS^2\over 2}
  \Bigl[ J \!\!\!\!\!\! \sum_{(i,j)\in n.N.} \!\!\!\!\!\! 
         \cos\alpha_i \cos\alpha_j
  -U_1 \!\!\!\!\!\! \sum_{(i,j)\in n.N.}  \!\!\!\!\!\! 
         \sin\alpha_i \sin\alpha_j 
  \nonumber \\
  && -2U_2  \!\!\!\!\!\! \sum_{(i,j)\in n.n.N.}  \!\!\!\!\!\!
        \sin\alpha_i \sin\alpha_j \Bigr]
     -\frac{1}{4}hS \sum_{i=1}^{4} \sin\alpha_i,
\label{Eg-4er}
\end{eqnarray}
where the sums run over the nearest neighbor ($\sum_{n.N.}$) and next-nearest
neighbor ($\sum_{n.n.N.}$) sites within the $2\times2$ unit cell.

\subsubsection{Two-sublattice supersolid phases: SS1 and SS1* phase}
\begin{figure}
\epsfxsize=0.7\columnwidth 
\centerline{\epsffile{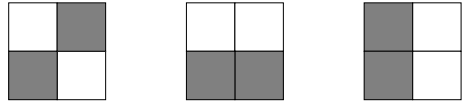}}
\vspace{0.3cm}
\caption{Sketch of the checkerboard and stripe symmetry. The left figure
corresponds to SS1 and the two right ones to SS1*. The four squares represent
the four spin angles $\alpha_1$-$\alpha_4$, where the same color means the same
angle. }
\label{fig:sketch}
\end{figure}

Starting from a general spin configuration with four different angles on a
$2\times2$ unit cell there are two different possibilities for a general
two-sublattice structure. Corresponding to the N\'eel and the stripe domain
zero-field structure one may have spin configurations with checkerboard
($\alpha_1=\alpha_3=\alpha$, $\alpha_2=\alpha_4=\beta$) and stripe symmetry
($\alpha_1=\alpha_2=\alpha$, $\alpha_3=\alpha_4=\beta$ or
$\alpha_1=\alpha_4=\alpha$, $\alpha_2=\alpha_3=\beta$), which we term SS1 and
SS1*, respectively (see Fig.~\ref{fig:sketch}).

Using such a two-sublattice Ansatz the corresponding ground state spin
configuration is obtained by minimizing the classical energy, Eq.~\ref{Eg-4er}.
This gives the two angles $\alpha$ and $\beta$ as a function of the magnetic
field $h$ and the interaction parameters. By comparing the resulting ground
state energy with the corresponding ground state energies of the N\'eel,
canted-ferromagnetic and $3/4$ phase upper and lower bounds for the stability
of the supersolid phase can be obtained. In order to show that these bounds are
actually phase boundaries, they must be compared with the results obtained
from a soft-mode analysis. Furthermore, it must be determined whether there are
additional intermediate phases with a spin structure more general than the
two-sublattice Ansatz discussed in this subsection.

After those general remarks, let us now discuss the explicit results from such
a mean-field analysis in the various parameter regions.

(i) For $U_1>2U_2$ the checkerboard structure is energetically preferred, 
i.e.\ we expect a phase transition from the N\'eel phase to the intermediate 
phase SS1. With the introduction of sum and difference angles
\begin{eqnarray}
  \gamma &=& \frac{\alpha + \beta}{2}, \\
  \delta &=& \frac{\alpha - \beta}{2}\, ,
\end{eqnarray}
minimization of the classical energy leads to a spin configuration described
by the angles
\begin{eqnarray}
\sin^2\gamma &=& {1\over 16SU_2}\sqrt{{U_1-U_2-J}\over U_1-U_2+J}(h-E_0), \\
\cos^2\delta &=& {1\over 16SU_2}\sqrt{{U_1-U_2+J}\over U_1-U_2-J}(h-E_0).
\label{equi-ss1}
\end{eqnarray}
The ground state energy reads
\begin{eqnarray}
E_g & = & -\frac{N}{32U_2}
 \left[ (h-E_0)^2+2(8S)^2U_2(U_1-U_2) \right],
\label{gs_energy_ss1}
\end{eqnarray}
where
\begin{eqnarray}
E_0 = 8S\sqrt{(U_1-U_2-J)(U_1-U_2+J)}.
\end{eqnarray}
Comparing with the ground state energy of the N\'eel phase,
Eq.~\ref{gs_energy_af}, one gets the lower bound of stability of the SS1 phase
\begin{equation}
  h_{\text{c1}} = E_0 = 8 S \sqrt{(U_1-U_2-J)(U_1-U_2+J)},
\end{equation}
which is identical to the upper stability boundary obtained from a soft-mode
analysis of the N\'eel phase, Eq.~\ref{af_critical_field}. Therefore, one may
conclude that this is the actual phase boundary between the N\'eel and SS1
phase, and the corresponding phase transition is continuous (at the mean-field
level).  The high-field boundary to the canted-ferromagnetic phase is obtained
by comparing with the corresponding classical ground state energy of the
canted-ferromagnetic phase, Eq.~\ref{gs_energy_canted}, or more conveniently by
the condition $\alpha = \beta$. One finds
\begin{equation}
h_{\text{c2}} = h_c \sqrt\frac{U_1-U_2-J}{U_1-U_2+J}, 
\end{equation}
which agrees with the lower stability bound obtained from a soft-mode analysis
of the canted-ferromagnetic phase, Eq.~\ref{h_c2}.  Hence, this is again an
actual phase boundary marking a continuous transition between the
canted-ferromagnetic and the supersolid SS1 phase. Of course, this conclusion
is valid only if the transition is not pre-emptied by a transition of the
canted-ferromagnetic phase to another phase which is lower in energy than the
SS1 phase. As already seen in the previous section there is a discontinuous
transition from $3/4$ phase to the canted-ferromagnetic phase for low $J$
values. The two phase-boundaries meet at the multicritical point mc$_1$
(s.\ Figs.~\ref{fig:pd_1_01} and \ref{fig:pd_1_02}) determined by
the implicit equation $h_{\text{c2}} = h_{\text{c3}}$.  Finally, by comparing
the classical ground state energies of the SS1 and the $3/4$ phase one
obtains the stability boundaries
\begin{eqnarray}
h^{\pm}_{\text{c4}} &=& E_0 +8SU_2 \nonumber \\
&&\pm\sqrt{16SU_2(E_0+4S(3U_2-2U_1))}.
\label{h_c4}
\end{eqnarray}
Due to different symmetries (two-sublattice for the SS1 and three-sublattice
for the 3/4 phase) this transition is discontinuous. This transition can appear
only for $J$ larger than the values at the multicritical point mc$_1$.  It
meets the phase boundaries $h_{\text{c2}}$ and $h_{\text{c3}}$ at mc$_1$ and
$h_{\text{c1}}$ at $J=0$ for a certain region of the parameters $U_1$ and
$U_2$.  Without a stability analysis of the SS1 phase, it is at this point not
possible to conclude that the line defined by Eq.~\ref{h_c4} is an actual phase
boundary. As we will see in the following sections the conclusion depends on
the actual choice of the parameters $U_1,U_2$ and $J$.

(ii) For $U_1<2U_2$ the low-field phase is the antiferromagnetic stripe phase.
The analog intermediate phase (SS1*) has a ground state energy of
\begin{equation}
E_g = -\frac{N}{16(U_1+J)}
\left[ (h-E_0^*)^2+(8S)^2U_2(U_1+J) \right] 
\end{equation}
with 
\begin{equation}
E_0^* =  8S\sqrt{U_2(U_2-J)}\, .
\end{equation}
Here the lower boundary, $h=E_o^*$, equals Eq.~\ref{af_critical_field} and
the upper boundary is
\begin{equation}
h_{c2}^* = h_c\sqrt\frac{U_2-J}{U_2}\, ,
\label{h_c2*}
\end{equation}
which agrees with Eq.~\ref{h_c21}. In the limit of small $J$ values this
transition is given by
\begin{equation}
h_{c2}^*\approx 8S(U_1+U_2) +4S{U_2-U_1\over U_2}J-{S(U_1+5U_2)\over
U_2^2}J^2\, . 
\label{h_c_ss1*_approx}
\end{equation}
Thus, the slope is positive for $U_2>U_1$, in contrast to all the other
transition lines at this point (except $h^{\text{para}}_{\text{min}}$). We will
see later that this influences the topology of the phase diagram.

The relation between the angles and the field is in analogy to the checkerboard
case 
\begin{eqnarray}
\sin^2\gamma &=& {1\over 8S(U_1+J)}\sqrt{{U_2-J}\over U_2}(h-E_0^*), \\
\cos^2\delta &=& {1\over 8S(U_1+J)}\sqrt{U_2\over U_2-J}(h-E_0^*).
\label{equi-ss1*}
\end{eqnarray}

Performing a stability analysis of both phases via linear spin-wave theory (see
appendix~\ref{app:sw_ss1}) shows the occurrence of another phase, a
three-sublattice phase SS2. The excitation spectrum for the SS1 phase has a
soft-mode at $q=\pi(1,0)$ and $q=\pi(0,1)$, whereas the SS1* has a soft-mode at
$q=\pi(1,1)$. The transition line has been determined numerically from the
expressions in the appendix. The soft-mode indicates a continuous transition.
However, the analysis of the SS2 phase in the next section shows that this is
only partially true.

\subsubsection{Three-sublattice supersolid phase: SS2 and SS2*}

\begin{figure}
  \epsfxsize=0.2\columnwidth \centerline{ \rotate[r]{
      \epsffile{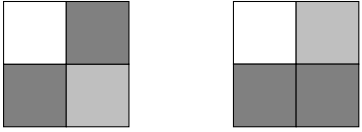}}}
    \vspace{0.3cm}
\caption{Sketch of the checkerboard and stripe symmetry. The left figure
  corresponds to SS2 and the right one to SS2* which is not stable in a
  four-sublattice model. The four squares represent the four spin angles
  $\alpha_1$-$\alpha_4$, where the same color means the same angle.}
\label{fig:sketch1}
\end{figure}

From spin-wave considerations in the last section we encountered another phase
which has to be more complicated than the general two-sublattice structure.
Therefore we are led to consider a three-sublattice structure (SS2 phase) and
calculate the phase boundaries to the other phases discussed above. We take the
following parameterization of the angles (s. Fig.~\ref{fig:sketch1})
\begin{equation}
\alpha_3=\alpha_1=\alpha,
\quad 
\alpha_2=\eta+\zeta,
\quad 
\alpha_4 = \eta-\zeta\, ,
\end{equation}
which has partially a checkerboard symmetry.  

Inserting in the general four-sublattice expression (Eq.~\ref{Eg-4er}) for the
ground state we can evaluate the mean-field equations for the angles and the
resulting ground state energy. Assuming a continuous transition from the SS2
phase in the $3/4$ phase we recover the result already found for the $3/4$
phase by spin-wave analysis, Eq.~\ref{sw_3/4}. This equation, as already
discussed, defines only partially a continuous transition.  Searching for a
continuous transition to the SS1 phase by comparing the corresponding ground
state energies we obtain the following equation
\begin{eqnarray}
hU_2E_0\sin{(\gamma-\delta)} =  &8S&[U_1(U_1-U_2)-J^2]h \nonumber \\
                                +&8S&(U_2^2-U_1^2+J^2)\, ,
\end{eqnarray}
which defines the same boundary as found numerically by the spin-wave analysis
of the SS1 phase.  The angles $\gamma$ and $\delta$ are given by the solution
for the SS1 phase (Eq.~\ref{equi-ss1}). Investigation of the ground state by
numerical minimization (simulated annealing, see next subsection) of the ground
state energy, Eq.~\ref{Eg-4er}, shows that there are regions where the
transition is continuous and regions where the transition is discontinuous. We
also find that there is no continuous transition to the SS1* phase possible due
to the different symmetry.

A corresponding SS2* spin structure with stripe symmetry can be parameterized
by (s. Fig. \ref{fig:sketch1})
\begin{equation}
\alpha_2=\alpha_1=\alpha,
\quad 
\alpha_3=\eta+\zeta,
\quad 
\alpha_4 = \eta-\zeta .
\end{equation}
Numerical investigation shows that this phase seems to be absent in the whole
parameter range. Within our numerical resolution we could not find any
significance for it. From symmetry considerations we expect that there is no
phase of that kind. There are only three different two-sublattice wave-vectors
$q_0,q_1$ and $q_2$ with which we cannot construct the symmetry of a SS2*
phase. 

\subsubsection{Four-sublattice supersolid phase: SS3}

In order to study the general four-sublattice structure we have employed two
different kinds of numerical methods. Using a simulated annealing code we have
determined the ground state spin configurations on the 2$\times$2 unit cell;
these simulations were crosschecked by a standard library for solving nonlinear
equations. We find that the general four-sublattice structure appears only for
$U_1<2U_2$, i.e. for the stripe phase ground state.  Thus we find an {\em
  intermediate supersolid phase (SS3)}, which interpolates between the SS1* and
SS2 phase such that both transitions become continuous. The resulting phase
diagrams are shown in the next section for various values of the interaction
parameters $U_1,U_2$ and $J$.

\section{Phase diagrams, multicritical points and universality classes}
\label{section:phase_diagram}

In this section we summarize our results for the phase diagram of the hard-core
Bose-Hubbard model with next-nearest neighbor interaction, as obtained from the
linear spin-wave analysis. Furthermore, we identify various multicritical
points and comment on the possible universality classes of the phase
transitions at the second order phase boundaries.

A number of different phase diagram topologies are possible depending on the
relative magnitude of the nearest and next-nearest neighbor interaction. Three
different classes may be distinguished. For small values of the next-nearest
neighbor interaction, $U_2 < U_1/2$, the zero-field ground state is
characterized by checkerboard symmetry. In an intermediate range, $U_1/2 < U_2
< U_1$, the zero-field ground state switches to stripe symmetry and there is
strong competition between nearest and next-nearest neighbor interaction.
Finally, the third class of phase diagram topologies is obtained when the
next-nearest neighbor interaction is dominant, $U_2>U_1$.

\subsection{Type I topology: 
  Checkerboard ground state and small next-nearest neighbor interaction ($\bf
  U_2 < U_1/2$)}

For relatively small next-nearest neighbor interactions, $U_2 < U_1/2$, the
classical ground state at zero magnetic field is the N\'eel state. If the
interaction is restricted to nearest neighbors ($U_2=0$) there are only three
phases: the N\'eel, the canted-ferromagnetic and the paramagnetic phase
        (s. Fig.~\ref{fig:pd_nearest}). Taking
into account longer-range interactions, i.e.\ for finite $U_2$, an additional
insulating phase with three quarter filling ($3/4$ phase) and two supersolid
phases, SS1 and SS2, appear \cite{bruder:93}.

\begin{figure}[htb]
  \centerline{\epsfxsize=0.75\columnwidth
    \epsffile{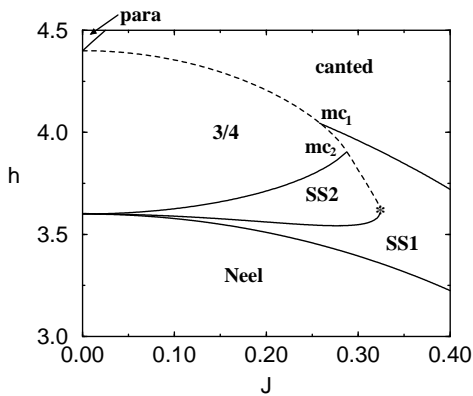}}
  \vspace{0.1cm}
\caption{Part of the phase diagram for the hard-core Bose-Hubbard model for 
  $U_1 = 1$ and $U_2 = 0.1$ (cf. \protect\cite{bruder:93}). Solid lines
  indicate continuous phase transitions (to mean-field level), whereas first
  order phase transitions are given by dashed lines. At finite $J$ there are
  four multicritical (mc) points: (i) two critical end points mc$_1$ and
  mc$_2$, (ii) a tricritical point on the tip of the supersolid SS2 phase
  (marked by a star (*)), and (iii) a multicritical point mc$_3$ on the tip of
  the N\'eel phase, which is not shown here.}
\label{fig:pd_1_01}
\end{figure}

\begin{figure}[htb]
  \centerline{\epsfxsize=0.75\columnwidth
    \epsffile{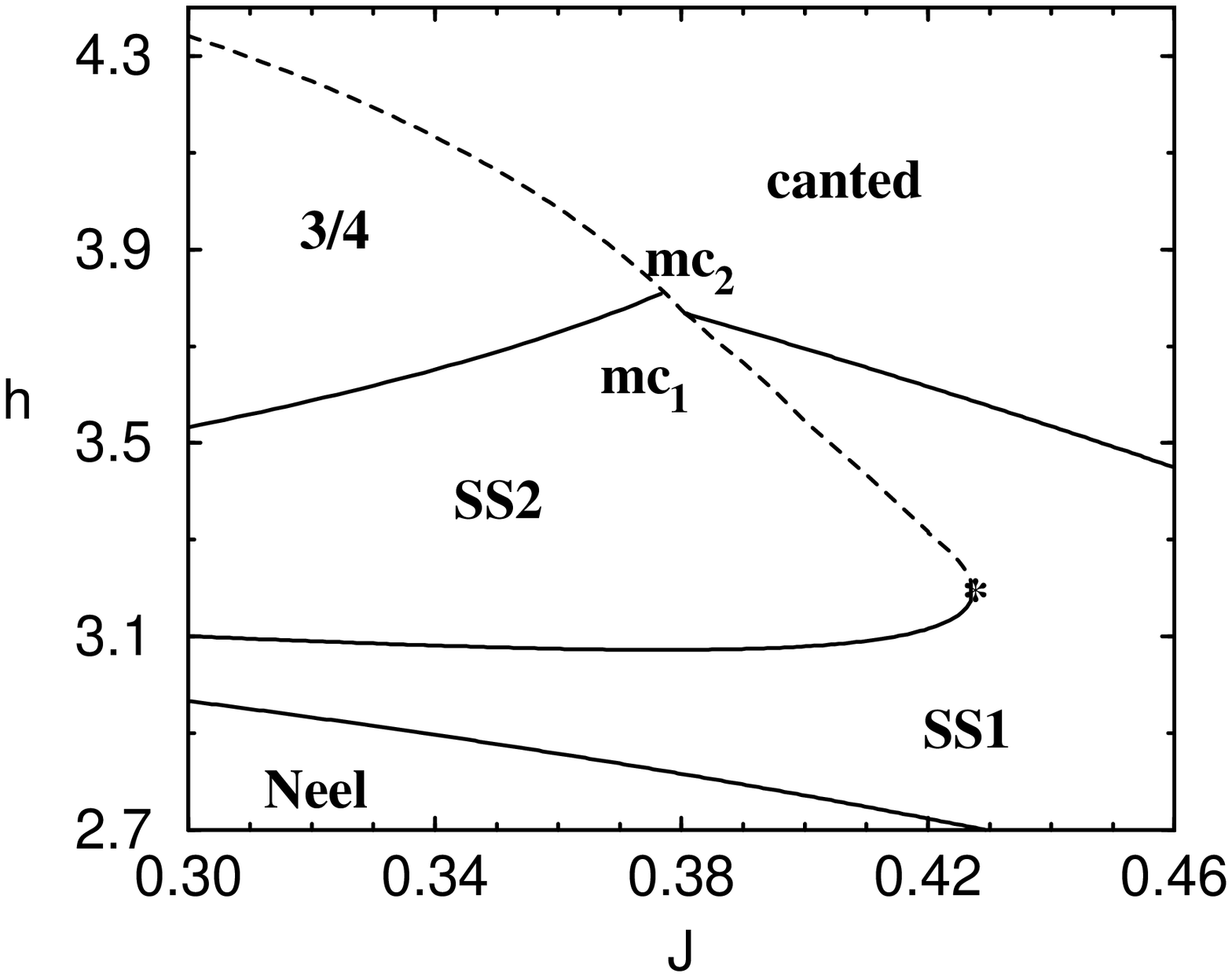}}
  \vspace{0.1cm}
\caption{The same as in Fig.~\protect\ref{fig:pd_1_01} for interaction
  parameters $U_1 =1$ and $U_2 = 0.2$.}
\label{fig:pd_1_02}
\end{figure}

\begin{figure}[htb]
  \centerline{\epsfxsize=0.75\columnwidth
    \epsffile{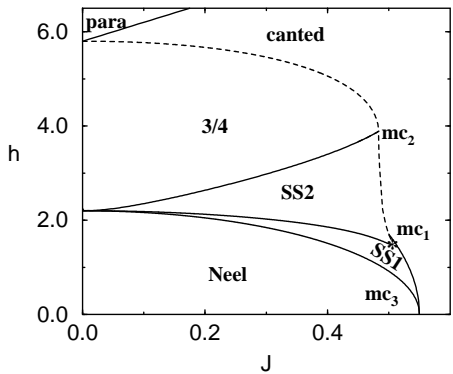}}
  \vspace{0.1cm}
\caption{The same as in Fig.~\protect\ref{fig:pd_1_01} for 
  interaction parameters $U_1 =1$ and $U_2 = 0.45$.}
\label{fig:pd_1_045}
\end{figure}

In Figs.~\ref{fig:pd_1_01}--\ref{fig:pd_1_045} we have shown the phase diagrams
resulting from ground-state calculations and linear spin-wave analysis for
$U_1=1$ and $U_2 = 0.1$, $U_2 = 0.2$, and $U_2 = 0.45$, respectively. The
characteristic features of the mean-field phase diagram are as follows.

(I) {\em Multicritical points:} At finite values of the hopping matrix element
there are four multicritical (mc) points mc$_1$, mc$_2$, mc$_3$ and the
tricritical point on the tip of the SS2 phase (marked by a star in
Figs.~\ref{fig:pd_1_01}-\ref{fig:pd_1_045}). Since for a finite value of the
next-nearest neighbor interaction $U_2$ there is an intermediate supersolid
phase intervening the N\'eel and the canted-ferromagnetic phase, there is a
multicritical point mc$_3$ at the tip of the N\'eel phase where two second
order phase boundaries meet each other with a common tangent.  For $U_2/U_1
\leq 0.183$ the second order line separating the canted-ferromagnetic (i.e.\ 
superfluid) from the supersolid SS1 phase intersects and is truncated by a
first order phase boundary to the commensurate $3/4$ solid phase. The
corresponding multicritical point mc$_1$ is a {\em quantum critical end point}.
The ``spectator phase'' is the uncritical Mott insulating 3/4 phase. The second
{\em quantum critical end point} mc$_2$ differs from mc$_1$ in one important
aspect. This time the ``spectator phase'' is the supersolid SS1 phase
characterized by the simultaneous presence of crystalline order and
superfluidity. Hence the ``spectator phase'' is critical since it shares the
superfluid order and the corresponding Goldstone mode for the phase of the
superfluid order parameter with one of the phases (the SS2 phase) associated
with the critical line. Fisher et al.~\cite{fisher-upton:90,fisher-barbosa:91}
have shown that the singular behavior at the critical end point also engenders
new singularities in the first order phase boundary itself.  Therefore one may
speculate that the presence of the critical end point may also affect the
``spectating'' supersolid phase and induce ``non Bose-liquid'' behavior.

Upon increasing the relative magnitude of the next-nearest neighbor interaction
the two critical end points mc$_1$ and mc$_2$ approach each other and for
$U_2/U_1 \approx 0.183$ they merge into a {\em higher order multicritical
  point}.  The critical values at this point are $J_c = 0.365$ and
$h_{\text{crit}} = 3.827$.  From a higher dimensional perspective, where the
ratio $U_2/U_1$ is added as an additional axis in the phase diagram, this point
corresponds to the crossing point of two multicritical lines. In the projection
of this higher dimensional phase diagram, mc$_1$ and mc$_2$ switch their
relative position for $U_2/U_1 > 0.183$: now at mc$_2$ the $3/4$, the
canted-ferromagnetic and the SS2, and at mc$_1$ canted-ferromagnetic, SS1 and
SS2 meet. 

There is also a tricritical point on the tip of the SS2 phase where the first
order line separating the two supersolid phases terminates and becomes second
order. The presence of the tricritical point is to be expected, since critical
end points in classical critical phenomena are intimately associated with the
vicinity of tricritical points \cite{fisher-barbosa:91}. With increasing the
next-nearest neighbor interaction the critical end point mc$_1$ and the
tricritical point approach each other. At $U_2 = U_1/2$ they merge into a
higher order multicritical point.

(II) {\em Supersolid phases:} There are two different supersolid phases, SS1
and SS2, and each of them has a Mott-insulating partner phase with the same
symmetry of the charge ordering, the N\'eel and the 3/4 phase, respectively.
The phase transition between the two supersolid phases is first-order at high
fields and becomes second-order at low fields. The tricritical point, where
those two transition lines meet is located at the tip of the SS2 phase. With
increasing $U_2$ the SS2 phase fills more and more of the parameter region
occupied by the SS1 phase (see Fig.~\ref{fig:pd_1_01}-\ref{fig:pd_1_045}). At
$U_1=2U_2$ the phase diagram changes drastically because the stripe phase
replaces the N\'eel phase.

\begin{figure}[htb]
  \centerline{\epsfxsize=0.6\columnwidth
    \epsffile{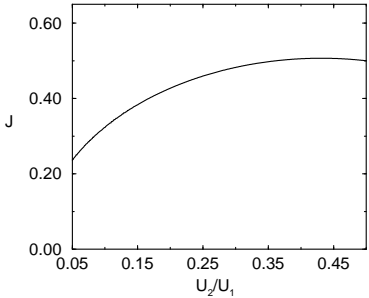}}
  \vspace{0.1cm}
\caption{Location of the tip of the supersolid SS2 phase as a function of 
  the ratio $U_2/U_1$. The maximum of the tip is reached before the transition
  to the stripe ground state, namely for $U_2/U_1\simeq 0.43$ with a maximum
  value of $J\simeq 0.507$.}
\label{fig:tip_ss2}
\end{figure}

In Fig.~\ref{fig:tip_ss2} a plot is shown for the tip ($J$-value, marked by a
star in the phase diagrams) of the SS2 lobe as a function of $U_2/U_1$. From
the soft-mode analysis (see appendix~\ref{app:sw_ss1}) of the SS1 phase we can
derive a cubic equation which defines the tip by the collapse of two solutions.
Surprisingly, the maximum of the tip is reached before the transition to the
stripe ground state, namely for $U_2/U_1\simeq 0.43$ with a maximum value of
$J\simeq 0.507$. As it turns out the SS1* phase intrudes in the area of the SS1
phase for $U_2$ slightly smaller than $0.5$ (s. Fig~\ref{fig:u2_inset}).

(III) {\em Universality classes:} Whereas the mean-field analysis in section
\ref{section:mean_field} is certainly sufficient in describing the overall
topology of the phase diagram and the ordered phases away from the transition,
it breaks down near the phase transition where the correlation length diverges.
The critical behavior at the phase transition from the SS1 phase (checkerboard
supersolid = XSS) to the canted-ferromagnetic phase (i.e.\ superfluid phase in
the corresponding Bose system) has recently been studied by renormalization
group theory \cite{frey-balents:97}. It is found that the phase transition from
the checkerboard supersolid to the superfluid phase can be described by an
effective action, where an Ising field is coupled to the superfluid phase
fluctuations ({\em XSS-SF universality class}). It exhibits nontrivial critical
behavior and appears within an $\varepsilon$-expansion to be driven first order
by fluctuations. However, within a calculation directly in $d=2$ dimensions one
finds a fixed-point with {\em ``non Bose-liquid''} behavior.  At the transition
between the two supersolid phases (at low fields) there are soft-modes at $q_1
= \pi (1,0)$ and $q_2 = \pi (0,1)$ (see appendix~\ref{app:sw_ss1}). Hence the
boson density in the SS2 phase may be characterized by
\begin{eqnarray}
n(x) = n_0 Re \left( 1 
      + \phi e^{i q_0 x} + \psi_1 e^{i q_1 x} + \psi_2 e^{i q_2 x} \right),
\end{eqnarray}
where $n_0$ is a background boson density and the fields $\phi$ and
$\bbox{\psi}$ represent the checkerboard and stripe domain order parameters,
respectively. From the mean-field analysis in the preceding section we know
that in the SS2 phase, the system orders along the $[11]$-direction in the
$(\psi_1,\psi_2)$-plane. This implies a particular form of the cubic anisotropy
in the effective action in the SS2 phase ($v' > 0$, see below). The XY field
$\bbox{\psi}$ is linearly coupled to the Ising field $\phi$ describing the
checkerboard order of the SS1 phase.  The corresponding action allowed by
symmetry reads \cite{frey-balents:97}
\begin{eqnarray}
  S_0 & = &
     \int d{\bf x} d\tau
     \Biggl\{
             \frac{1}{2}\left( \frac{1}{c}
             \partial_\tau \phi \right)^2 +
             \frac{1}{2}\left( \bbox{\nabla} \phi \right)^2
             + \frac{t}{2} \phi^2 + \frac{u}{4!} \phi^4 \nonumber \\
             && + \frac{1}{2}\left| \frac{1}{{c'}}
             \partial_\tau \bbox{\psi} \right|^2 +
             \frac{1}{2}\left| \bbox{\nabla} \bbox{\psi} \right|^2
             + \frac{t'}{2}|\bbox{\psi}|^2
             + \frac{u'}{4!}|\bbox{\psi}|^4\nonumber \\
             &&+ \frac{v'}{4!} \sum_i \psi_i^4 
               + w \phi \prod_i \psi_i
               + {\tilde w} \phi^2 \sum_i \psi_i^2
     \Biggr\} \, ,
\label{critical_action}
\end{eqnarray}
with the control parameters $t$ and $t'$ measuring the distances from the
critical lines. Within the linear spin-wave theory in section
\ref{section:mean_field} these critical lines coincide with those curves in
parameter space, where the spin-wave spectrum has a soft-mode at the ordering
wave vector of the corresponding field. The parameters of the effective action
are related to the microscopic parameters of the original Bose-Hubbard model,
which may be worked out using a mean-field decoupling procedures (see e.g.\ the
appendix of Ref.~\cite{frey-balents:97}). Non-local interactions arise due to
interactions with long wavelength fluctuations of the superfluid phase,
$\theta$,
\begin{equation}
 S_1 = \frac{\tilde{\rho}_s}{{2m^2}}\int \! d{\bf x}d\tau \bigg\{
            \left(\frac{1}{v} \partial_\tau\theta\right)^2 +
            (\bbox{\nabla}\theta)^2\bigg\},
\label{phase_action}
\end{equation}
where we have used the same notation as in Ref.~\cite{frey-balents:97}.  This
part of the effective action originates from the $x$ and $y$ spin components in
the spin-$1/2$ XXZ model, Eq.~\ref{XXZ}. Sufficiently far from the phase
boundary where the superfluid order parameter vanishes, the most relevant
coupling to the spatial order parameters allowed by the time-reversal and
$U(1)$ symmetries is \cite{frey-balents:97}
\begin{equation}
  S_2 =  \int \! d{\bf x}d\tau 
         \bigg\{  i \sigma \partial_\tau\theta
                      |\phi|^2 + i \sigma' \partial_\tau\theta
                      |\bbox{\psi}|^2 \bigg\}.
\label{coupling_action}
\end{equation}
For the phase transition from the SS1 to the SS2 phase, the spin-wave spectrum
softens at the stripe domain ordering wave vectors $q_1$ and $q_2$, i.e.\ $t'$
becomes zero. Since this happens in the presence of an already ordered Ising
field $\phi$ there are now two different cubic anisotropies, the quartic term
and the quadratic term, $g \psi_1 \psi_2$, where $g = w \langle \phi \rangle$.
Whereas the quartic cubic anisotropy is an irrelevant symmetry breaking field
for the XY transition~\cite{note}, the quadratic symmetry breaking field $g$
leads to an uniaxial anisotropy in the $[11]$-direction and hence to a
reduction in the number of soft spin components from $n=2$ to $n=1$. This in
turn implies that the coupling to the phase fluctuations becomes relevant to
the asymptotic critical behavior at the SS1 to SS2 phase transition, and the
asymptotic critical behavior is in the XSS-SF universality class. When the
checkerboard order parameter $\langle \phi \rangle$ is decreased, the quadratic
symmetry breaking field becomes small, and there can be quite interesting
crossover phenomena. One should note that here both the cubic and the quartic
cubic anisotropy favor spin orientation in the $[11]$-direction, since $v' >
0$. Later we will encounter a case where the cubic anisotropy becomes negative
and as a consequence of the competition between these two anisotropies new
intermediate supersolid phases appear.

Interesting new quantum critical behavior is also found in the vicinity of the
tricritical point at the tip of the supersolid SS2 phase. Since the tricritical
point of the classical $D=3$ $n$-vector model is described by classical
exponents with logarithmic corrections, the specific heat exponent $\alpha$
becomes positive, $\alpha = 2 - D \nu = 0.5$. As a consequence the coupling to
superfluid phase mode becomes even more relevant on approaching the tricritical
point. The effective action at the tricritical point reduces to
\begin{eqnarray}
  S_0 & = &
     \int d{\bf x} d\tau
     \Biggl\{\frac{1}{2}\left| \frac{1}{c'}
             \partial_\tau \bbox{\psi} \right|^2 +
             \frac{1}{2}\left| \bbox{\nabla} \bbox{\psi} \right|^2
             + \frac{t'}{2}|\bbox{\psi}|^2 + \frac{v_1'}{6!}|\bbox{\psi}|^6
             \nonumber \\
             &&+ \frac{v_2'}{6!}|\bbox{\psi}|^2  \sum_i \psi_i^4 + 
                 \frac{v_3'}{6!} \sum_i \psi_i^6 + g \psi_1 \psi_2 \Biggr\}.
\label{tricritical_action}
\end{eqnarray}
Due to the positive value of $\alpha$, a renormalization group analysis of this
{\em quantum tricritical model} will yield a new universality class different
from the $D=3$ tricritical Ising model. We leave the corresponding analysis of
the asymptotic critical and crossover behavior for a future investigation
\cite{frey-pich:97}.

Finally, the topology of the phase diagram allows for phase transitions between
the supersolid and commensurate solid phases. At both transition the order
parameter characterizing superfluid order becomes zero. In passing from the SS1
to the N\'eel phase the order parameter describing checkerboard ordering
changes smoothly, while a two-component field $\Psi$ describing superfluidity
becomes critical. Since generically there is no particle-hole symmetry (charge
conjugation) symmetry along this phase boundary
\cite{fisher-weichman-grinstein:89}, the effective action can contain a term
$\Psi^\star \partial_\tau \Psi$ which is more relevant than $|\partial_\tau
\Psi|^2$. Hence, the transition in $2+1$ dimensions becomes mean-field like
with logarithmic corrections. Similar conclusions hold for the transition from
the SS2 to the 3/4 phase.

\subsection{Type II topology: 
  Intermediate next-nearest neighbor interaction ($\bf U_1/2< U_2 < U_1$)}

In this parameter region the zero-field ground state is an antiferromagnet with
a striped structure. As a consequence of the strong competition between nearest
and next-nearest neighbor interaction the phase diagram exhibits quite a rich
structure with first and second order phase boundaries and various
multicritical points. The characteristic features are as follows:

\begin{figure}[htb]
  \centerline {\epsfxsize=0.75\columnwidth
    \epsffile{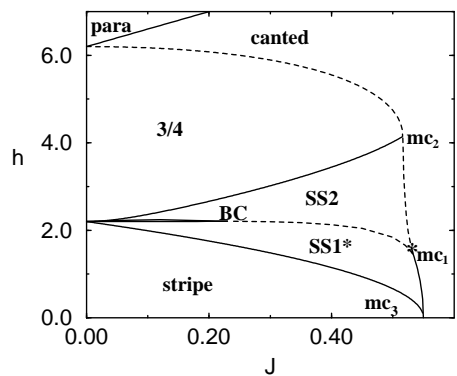}}
  \vspace{0.1cm}
\caption{Phase diagram for the hard-core Bose-Hubbard model with interaction
  parameters $U_1 = 1$ and $U_2 = 0.55$. The first-order and second-order lines
  in the phase diagram are depicted by dashed and solid lines, respectively.
  There are five multicritical points: (i) one critical end point, mc$_2$, (ii)
  two bicritical points, BC and mc$_1$, (iii) one tricritical point along the
  phase boundary between the supersolid SS2 phase and the canted-ferromagnetic
  phase in the intermediate vicinity of mc$_1$ (marked by a star (*)), and (iv)
  the multicritical point on the tip of the checkerboard solid phase. The small
  sliver is the SS3 phase.}
\label{pd_1_055}
\end{figure}

\begin{figure}[htb]
  \centerline {\epsfxsize=0.75\columnwidth
    \epsffile{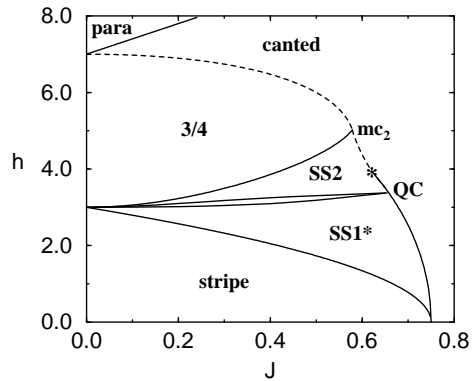}}
  \vspace{0.1cm}
\caption{Phase diagram for the hard-core Bose-Hubbard model with interaction
  parameters $U_1 = 1$ and $U_2 = 0.75$. The bicritical point on the tip of the
  SS3 phase has merged with the bicritical mc$_1$ to from a tetracritical point
  (QC). Otherwise the phase diagram has the same topology as
  Fig.~\ref{pd_1_055}.}
\label{pd_1_075}
\end{figure}
\begin{figure}[htb]
  \centerline {\epsfxsize=0.75\columnwidth
    \epsffile{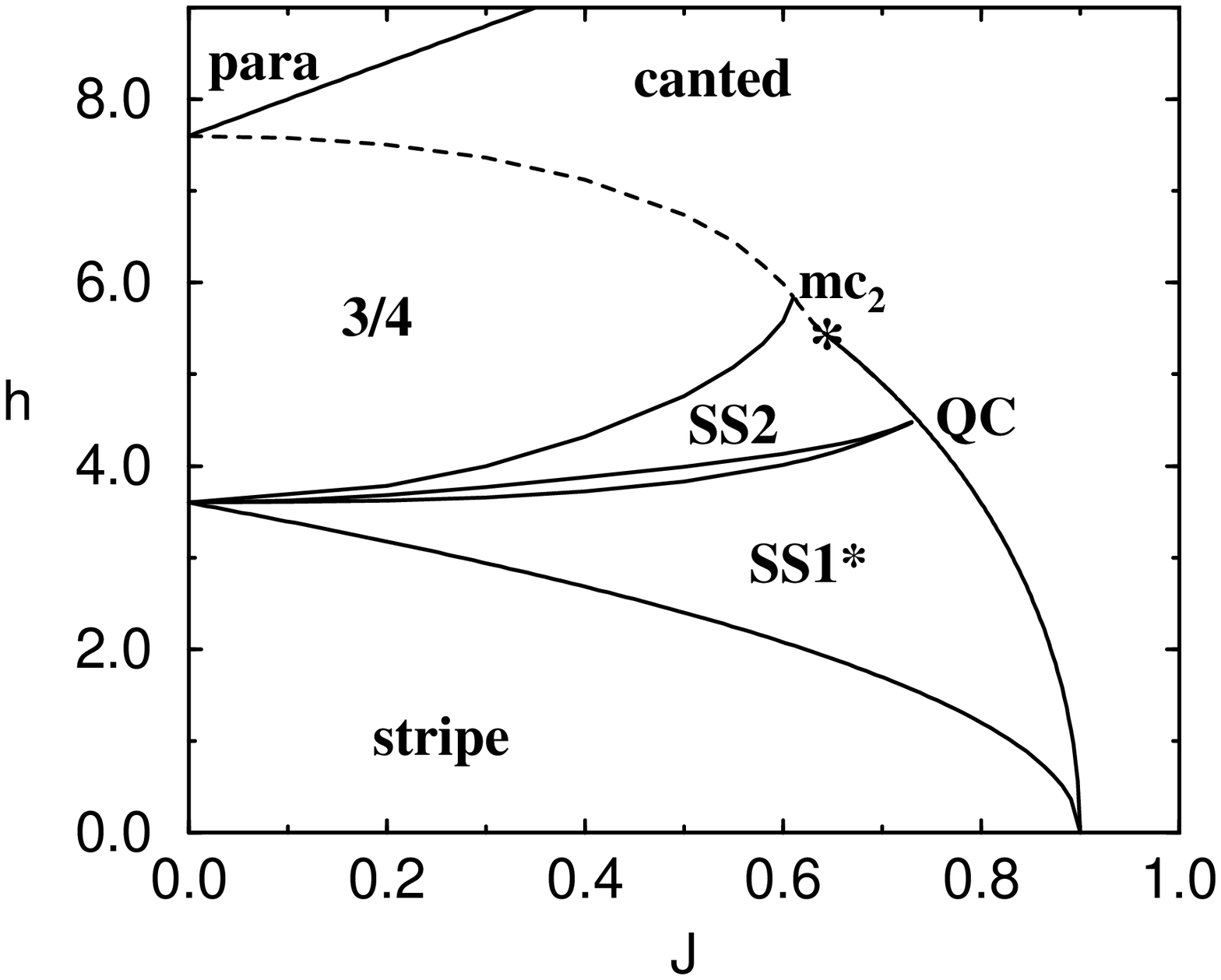}}
  \vspace{0.1cm}
\caption{Phase diagram for the hard-core Bose-Hubbard model 
  with interaction parameters $U_1 = 1$ and $U_2 = 0.9$. The phase diagram has
  the same topology as Fig.~\protect \ref{pd_1_075}. The tricritical point on
  the phase boundary between the SS2 and the canted-ferromagnetic phase has
  shifted towards the multicritical point mc$_2$.}
\label{fig:pd_1_09}
\end{figure}

\begin{figure}[htb]
  \centerline {\rotate[r]{\epsfxsize=0.7\columnwidth
  \epsffile{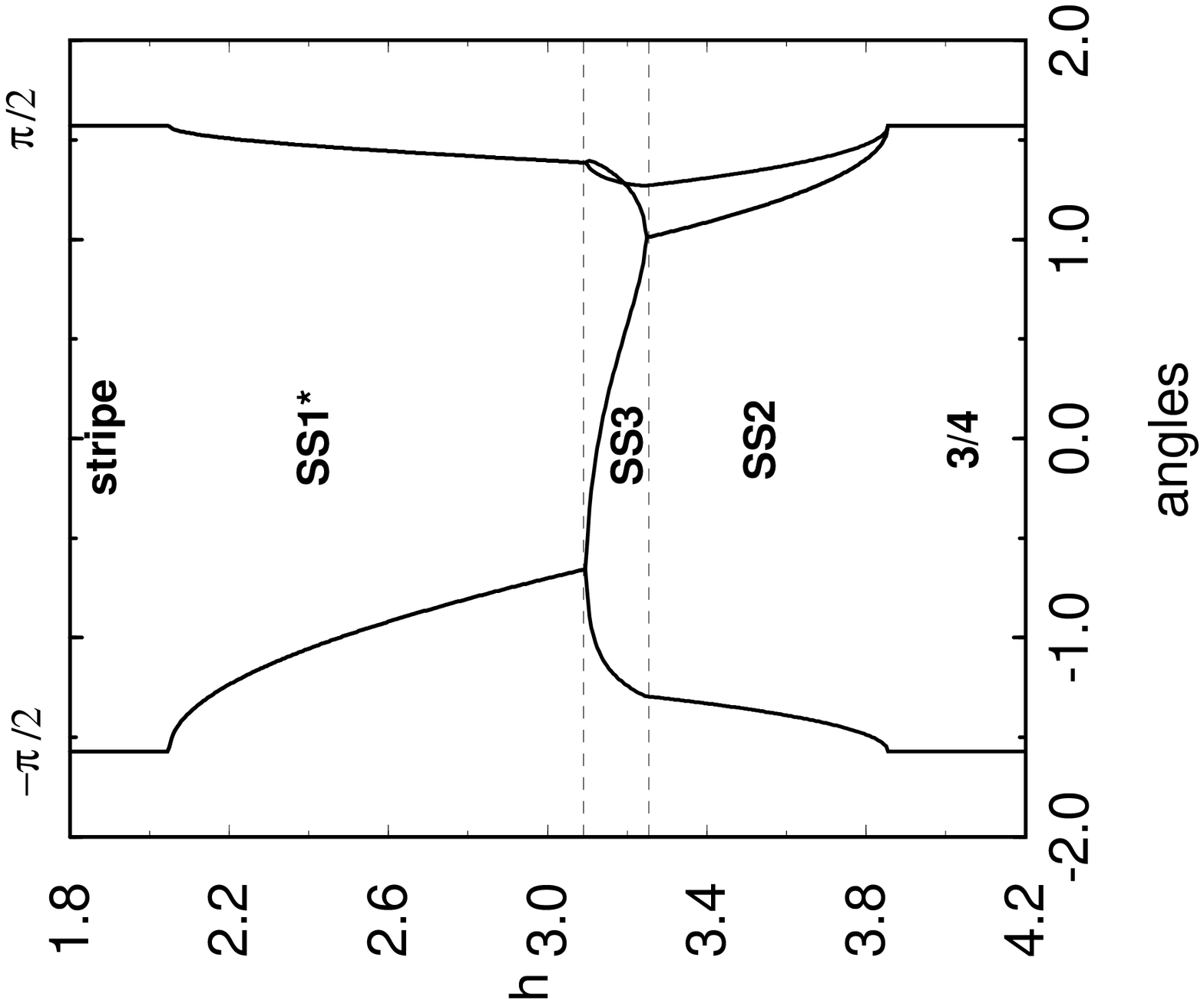}}}
  \vspace{0.5cm}
\caption{Variation of the angles on the $2\times2$ unit cell characterizing 
  the spin configuration at fixed interaction parameters $J=0.40$, $U_1= 1$ and
  $U_2 = 0.75$ as a function of the magnetic field $h$. All transitions are
  seen to be continuous on the mean-field level. The boundaries of the SS3
  supersolid phase are marked by dashed lines.}
\label{angles_1_075}
\end{figure}

(I) {\em Supersolid phases:} There are {\em three} supersolid phases. In
addition to the SS1 and SS2 phase we find an intermediate supersolid phase SS3
characterized by a four-sublattice structure with all four angles on the
$2\times2$ unit cell being different. Increasing the next-nearest neighbor
interaction $U_2$ from the lower bound $U_2/U_1=0.5$ a sliver of the SS3 phase
appears at small values of the hopping matrix element $J$ (see
Fig.~\ref{pd_1_055}). Both of the transitions of the other two supersolid
phases to this intermediate supersolid phase are continuous. This can also be
inferred from Fig.~\ref{angles_1_075}, where the variation of the four angles
of the spin configuration are shown as a function of the magnetic field $h$ for
fixed hopping matrix element $J=0.40$. 
For larger values of $J$ the transition is still directly from the SS2 to the
SS1* phase with a jump in the angles of the spin configuration, i.e.\ the
transition is first-order. As the ratio of $U_2/U_1$ becomes larger the SS3
phase occupies a larger portion in parameter space and finally (at $U_2/U_1
\approx 0.75$) meets with its tip the phase boundary to the
canted-ferromagnetic phase. See the topology of the phase diagram in
Fig.~\ref{pd_1_075}, where $U_2/U_1 = 0.75$.  For larger values of $U_2$ the
SS3 phase keeps its topology, i.e.\ there is an intermediate SS3 phase
completely separating the SS2 from the SS1* phase by a series of continuous
phase transitions.

(II) {\em Multicritical points:} Due to the competition between nearest
neighbor and next-nearest neighbor interaction there are also several
interesting multicritical points. Here we discuss those which appear at finite
values of $J$. Right at $U_2 = U_1/2$ the critical end point mc$_1$ and the
tricritical point at the tip of the SS2 phase (compare Fig.~\ref{fig:pd_1_045})
have merged into a {\em tricritical end point}. A slight increase of $U_2$
above the critical value $U_1/2$ implies that part of the phase boundary
between the SS2 and the canted-ferromagnetic phase becomes a second order line.
Then, mc$_1$ becomes a bicritical point and a new tricritical point close to
mc$_1$ appears (compare Fig.~\ref{pd_1_055}). The bicritical point separates
two second-order lines from the superfluid to two supersolid phases with
distinct ordering wave vectors. The SS1* phase is characterized by one of the
stripe-domain ordering wave vectors $q_1 = \pi (0,1)$ or $q_2 = \pi (1,0)$;
the charge ordering in the SS2 phase is more complicated and in addition to
$q_1$ and $q_2$ requires a third (checkerboard) wave vector $q_0 = \pi
(1,1)$.  In addition there is a bicritical point (BC) on the tip of the SS3
phase. Upon increasing the next-nearest neighbor interaction the tip of the SS3
phase and hence the bicritical point moves towards the bicritical point mc$_1$
until both join at the critical value $U_2^\star \approx 0.75$ to a
tetracritical point (QC) \cite{note_tetra}. At the same time the phase boundary
connecting mc$_1$ and mc$_2$ becomes more and more of second order.  As a
consequence the tricritical point shifts toward the critical end point mc$_2$
(compare Fig.~\ref{fig:pd_1_09}). Finally, at $U_2 = U_1$ the phase boundary
between the SS2 and canted-ferromagnetic phase is completely of second order
and the tricritical point merges with the critical end point to become a
tricritical end point at
\begin{equation}
J_c = g-1\, , \qquad h_c = 4 g\, ,
\end{equation}
where $g=(1+\sqrt 5)/2$ denotes the golden mean. The tip of the $3/4$ phase is
located at $J_{\text{tip}} = (2/\sqrt{27})^{1/2}>J_c$. This tip appears for
$U_2 > 0.96$. At $U_2=U_1$ the phase line for smaller $J$ values is still a
discontinuous transition between the canted-ferromagnetic and the $3/4$ phase
but becomes a complex transition for slightly larger values (s.\ subsection C).

(III) {\em Universality classes:} In each of the three supersolid phases the
boson density may again be characterized by 
\begin{eqnarray}
n(x) = n_0 Re \left( 1 
           + \phi e^{i q_0 x} + \psi_1 e^{i q_1 x} + \psi_2 e^{i q_2 x} 
              \right).
\end{eqnarray}

For $U_2/U_1 \leq 0.75$ and large enough $J$ there is a first order line
separating the SS1* from the SS2 phase. In both phases the stripe domain order
parameter $\bbox{\psi}$ is finite, but differs in its alignment in the
$(\psi_1,\psi_2)$-plane.  Whereas the vector $\bbox{\psi}$ points along one of
the cube edges in the SS1* phase, it is oriented along one of the cube
diagonals in the SS2 phase. From the mean-field analysis in section
\ref{section:mean_field} we also know that the checkerboard order parameter
$\phi$ is zero in the SS1* phase but finite in the SS2 phase. Below a certain
critical value of $J$ which depends on the ratio $U_2/U_1$ there is an
intervening supersolid phase (SS3) characterized by an order parameter
$\bbox{\psi}$ which rotates from the cube edges to the cube diagonals as one
passes from the SS1* to the SS2 phase.

The topology of this part of the phase diagram can be understood in terms of a
mean-field analysis of the effective action, Eq.~\ref{critical_action}. The
phase boundary between the SS1* phase and the SS3 phase corresponds to $t=0$
and the phase boundary between the superfluid phase and both of the supersolid
phases corresponds to $t'=0$. The first order transition between the SS1* and
the SS2 phase is a consequence of the competition between the quartic ($v'<0$)
and quadratic ($w$) cubic symmetry breaking fields in the effective action.  A
mean-field analysis of the latter shows that sufficiently far from the phase
boundary $t=0$ the first order line is given by the relation $v'_{c} = -6 w^2 /
t$.  For $v' < v_c$ the effective action is minimized by a configuration, where
$\phi = 0$ and $\bbox{\psi}$ points along one of the cube edges. This is the
SS1* phase. For $v' < v_c$ both the checkerboard and stripe domain order
parameters become finite and $\bbox{\psi}$ points along one of the cube
diagonals, as is the case for the SS2 phase. In particular one finds that at
the phase transition from the superfluid to the SS2 phase ordering of the
$\bbox{\psi}$-field induces ordering in the Ising field $\phi$. In mean-field
(mf) approximation one gets $\phi_{\text{mf}} = (-w) \langle \psi \rangle^2 /
t$, due to the trilinear coupling to the $\bbox{\psi}$-field.

The competition between the quartic and quadratic cubic symmetry breaking
fields also leads to the existence of an intermediate supersolid phase (SS3).
In the SS1* phase the quartic field $v' < 0$ dominates and we have alignment
along one of the cube edges in the $(\psi_1,\psi_2)$-plane. Right at the phase
boundary between the SS1* and the SS3 phase the spin-wave spectrum becomes soft
at the checkerboard ordering wave vector $q_0$ (i.e.\ $t=0$). Hence within the
SS3 phase we have a finite mean value for the Ising field, $\langle \phi
\rangle \neq 0$, which implies that in addition to the quartic cubic anisotropy
($v'<0$) we have a quadratic term, $ g \psi_1 \psi_2 $ with $g = w \langle \phi
\rangle$.  The latter favors spin alignment along the cube diagonals in the
$(\psi_1,\psi_2)$-plane, which now starts competing with the quartic symmetry
breaking field $v' < 0$. By increasing the magnetic field and moving away from
the SS1*-SS3 phase boundary the expectation value of the Ising field $\phi$ and
hence the magnitude of the quadratic symmetry field $g$ starts to grow leading
to a rotation of the spin alignment from the cube edges to the cube diagonals
(compare also Fig.~\ref{angles_1_075}). When the rotation is completed there is
a phase transition into the SS2 phase. With increasing $U_2/U_1$ the bicritical
point on the tip of the SS3 phase meets with the phase boundary to the
canted-ferromagnetic phase and it becomes a tetracritical point
\cite{note_tetra}.

The above symmetry considerations allow us now to discuss the universality
classes of the phase transitions between the various supersolid phases and the
superfluid phase. Let us first discuss the phase transitions between the
superfluid and the supersolid phases. The transition from the superfluid to the
SS1* phase (collinear supersolid = CSS) is in the universality class of the
$D=3$ classical XY model ({\em CSS-SF universality class})
\cite{frey-balents:97}. At the second order phase boundary between the
superfluid and the SS2 phase (mixed supersolid = MSS) the checkerboard and the
stripe domain ordering wave vectors become soft simultaneously. As discussed
above, this is not a consequence of both $t$ and $t'$ becoming zero at this
phase boundary but due to the trilinear coupling between the checkerboard and
stripe domain order parameters. Ordering of the stripe domain field
$\bbox{\psi}$ automatically induce ordering of the checkerboard field $\phi$
(see the above mean-field result).  The critical exponent $\beta_\phi$ for the
Ising field is $\beta_\phi = 1$ already at the mean-field level. Whereas the
critical behavior of the XF field seems to be unchanged at the transition, the
trilinear coupling induces cusp-like singularities in the {\em ``slaved''
checkerboard field} $\phi$. The critical behavior of such a model consitutes a
new universlity class {\em (MSS-SF universality class}) and leave a more
detailed analysis for a future investigation \cite{frey-pich:97}.
  
Next we consider the universality classes of the phase transitions between the
various supersolid phases. By increasing the field $h$ in the SS1* phase one
may either encounter a first order phase boundary to the SS2 phase or a second
order boundary to the SS3 phase. In the latter case the spin-wave spectrum
becomes soft at the checkerboard ordering wave vector. Thus the effective
action is given by an Ising-field which couples linearly to an XY field which
is already ordered along one of the cube edges. After integrating out the
$\bbox{\psi}$ field one is left with an Ising model with renormalized
coefficients. Hence this transition belongs to the XSS-SF universality class
\cite{frey-balents:97}. In the intermediate supersolid SS3 phase the direction
of the XY field $\bbox{\psi}$ rotates from the cube edge to the cube diagonal.
The order parameter for the transition to the SS2 phase is the one-component
field $\sigma = \psi_1 - \psi_2$. We thus conclude that this transition is
again the XSS-SF universality class.

At the multicritical point on the tip of the SS3 phase both of the Ising fields
$\phi$ and $\sigma = \psi_1 - \psi_2$ become critical.  The corresponding
{\em bicritical} effective action for the charge ordering fields is given
by
\begin{eqnarray}
  S_0^{\text{bc}} & = &
     \int d{\bf x} d\tau
     \Biggl\{\frac{1}{2}\left( \frac{1}{c}
             \partial_\tau \phi \right)^2 +
             \frac{1}{2}\left( \bbox{\nabla} \phi \right)^2
             + \frac{t}{2} \phi^2 + \frac{u}{4!} \phi^4 \nonumber \\
             && +  \left( \frac{1}{c'} \partial_\tau \sigma \right)^2 +
             \frac{1}{2}\left( \bbox{\nabla} \sigma \right)^2
             + \frac{t}{2} \sigma^2 + \frac{u'}{4!} \sigma^4 \nonumber \\
             && + w \phi \sigma^2 \Biggr\} \, .
\label{bicritical_action}
\end{eqnarray}
Note that this is a rather peculiar effective action for a bicritical point,
since it consists of two Ising fields $\phi$ and $\sigma$ coupled by a
three-point vertex. Since the upper critical dimension of the coupling vertex
$w \phi \sigma^2$ is $D_c=6$ as compared to $D_c=4$ for the $\phi^4$ and
$\sigma^4$ vertices the former will dominate the critical behavior.  In
combination with the coupling to the superfluid phase this should result in
quite interesting critical behavior.  The effective action bears some
resemblance with field theories for anisotropic Potts models
\cite{theumann-theumann:82}.  It would, however, go beyond the scope of the
present paper to analyze the critical behavior of such a model.  But, it
certainly constitutes a quite interesting new {\em quantum bicritical
  universality class}, which could be analyzed using standard renormalization
group theory.

With increasing the next-nearest neighbor interaction $U_2$ the bicritical
point moves towards the phase boundary to the superfluid phase. When it meets
at about $U_2/U_1 = 0.75$ one gets a tetracritical point, where four second
order lines meet. 

Finally, the phase transition from the supersolid SS1* phase to the stripe
domain solid and from the SS2 phase to the 3/4 phase are both most likely
mean-field like with logarithmic corrections in $D=2+1$ dimensions.

\subsection{Type III topology: 
  Large next-nearest neighbor interaction ($\bf U_2 > U_1$)}

For this parameter range the zero-field classical ground state is an
antiferromagnet with stripe symmetry. Due to the positive slope of the critical
field in the $SS1*$ phase (Eq.~\ref{h_c_ss1*_approx}) the SS2 phase transition
splits the discontinuous transition from the canted-ferromagnetic to $3/4$
phase into two continuous transitions. For $U_2> 1.149$ this transition is
established for the whole high field region.

\begin{figure}[htb]
  \centerline {\epsfxsize=0.75\columnwidth
    \epsffile{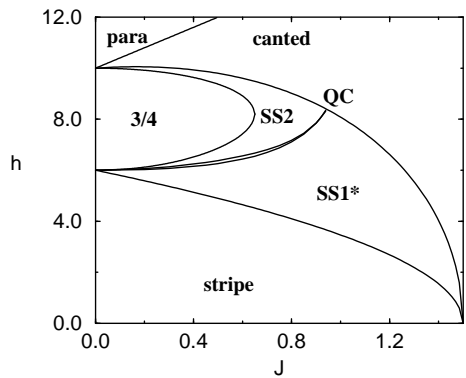}}
  \vspace{0.1cm}
\caption{Phase diagram for the hard-core Bose-Hubbard model with interaction
  parameters $U_1 = 1$ and $U_2 = 1.5$. The phase in between the SS2 and the
  SS1* is the SS3.}
\label{fig:pd_1_15}
\end{figure}

{\em (I) Supersolid phases and topology of the phase diagram:} For values of
the next-nearest neighbor interaction in the range of $1.149< U_2/U_1 \lesssim
3$ (see Fig.~\ref{fig:pd_1_15}) 
there are three different supersolid phases, SS1*, SS2, SS3. The SS3 phase is
located in the thin sliver between the SS2 and SS1* phase. Due to the presence
of the intermediate SS3 supersolid phases all phase transitions are continuous.
The transition from the SS2 phase directly to the canted-ferromagnetic phase
without passing the SS1* phase has also quite interesting properties.
Surprisingly, the transition line is given by
\begin{equation}
h = h_c\sqrt{{U_2-J}\over U_2},
\label{cant-ss2}
\end{equation}
the same dependence as for the transition of canted-ferromagnet to SS1*. 

\begin{figure}[htb]
  \centerline {\epsfxsize=0.75\columnwidth
    \epsffile{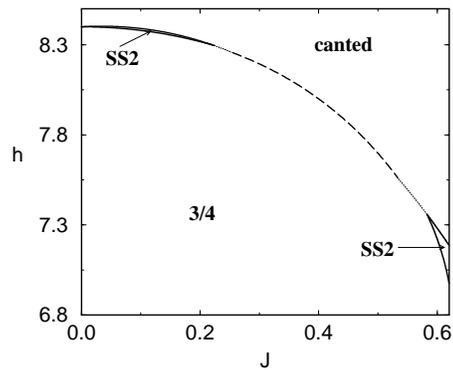}}
  \vspace{0.1cm}
\caption{Phase diagram for the hard-core Bose-Hubbard model with interaction
  parameters $U_1 = 1$ and $U_2 = 1.1$. The dashed transition line defines a
  discontinuous transition from canted-ferromagnetic to 3/4, the solid lines
  for small J denotes two second order transition, namely from
  canted-ferromagnetic to SS2 and then to 3/4. The dotted lines denote a
  discontinuous transition from canted-ferromagnetic to SS2 and a continuous
  transition from SS2 to 3/4. The actual phase cannot be resolved on this
  scale. }
\label{fig:pd_1_11}
\end{figure}

The topology for $1< U_2/U_1 < 1.149$ is similar to the just described one but
with a slight change for the high field transition between the
canted-ferromagnetic and the $3/4$ phase. When comparing the phase transition
between the canted-ferromagnetic and the $3/4$ (Eq. \ref{cant-34}), the
canted-ferromagnetic and the SS2 (Eq. \ref{cant-ss2}) and the $3/4$ and the SS2
(Eq. \ref{sw_3/4}) we encounter various region (as a function of $J$) where the
SS2 phase occurs. In the region for $U_2/U_1 < 1.124$ we have for increasing
$J$ the following picture (Fig.~\ref{fig:pd_1_11}): There is a SS2 phase
between the canted and the $3/4$ (continuous on both transitions), then a SS2
phase with a discontinuous transition to the $3/4$ phase, a direct transition
from canted-ferromagnetic to $3/4$, a SS2 with discontinuous transition to the
$3/4$ again and for high values of $J$ the SS2 phase with continuous
transitions to either phases.
The direct transition from canted-ferromagnetic to $3/4$ vanishes for $U_2/U_1
\geq 1.124$ at $J = 0.415$. For $U_2/U_1>1.149$ only the SS2 phase with the
continuous transitions keeps established. This happens for a critical value of
$J=0.45$. Equality of Eq. \ref{h_c2*} and Eq.  \ref{h_c3} define those
multicritical points where the canted to $3/4$ and the canted to SS2 phase
lines meet. They are given by the two positive solutions of (besides $J=0$)
\begin{eqnarray*}
J^3+(2U_1-U_2)J^2+U_1(U_1-2U_2)J+U_2(U_2^2-U_1^2) =0\, .
\end{eqnarray*}

\begin{figure}[htb]
  \centerline{\epsfxsize=0.7\columnwidth
    \epsffile{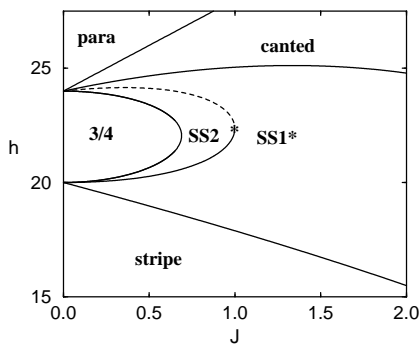}}
  \vspace{0.1cm}
\caption{Phase diagram for the hard-core Bose-Hubbard model for 
 $U_1 = 1$ and  $U_2 = 5$. The SS3 phase, which is located between the SS2 and
 the SS1* (lower fields) cannot be resolved at this scale.}
\label{pd_02_1}
\end{figure}

\begin{figure}[htb]
  \centerline{\epsfxsize=0.7\columnwidth
    \epsffile{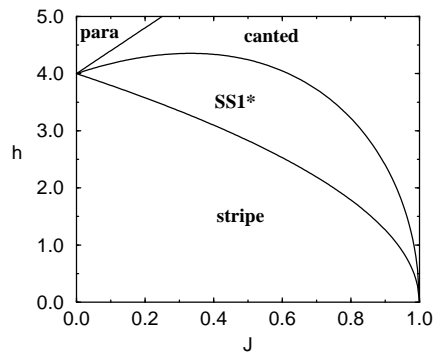}}
  \vspace{0.1cm}
\caption{Phase diagram for the hard-core Bose-Hubbard model for 
  pure next-nearest interaction, $U_2=1$ and $U_1=0$. Note that there is only
  one supersolid phase (SS1*).}
\label{pd_0_1}
\end{figure}

For large values of next-nearest neighbor interaction, $U_2/U_1>3.1$, the phase
diagram has a topology as depicted in Fig.~\ref{pd_02_1} for the particular
value of $U_1 = 1$ and $U_2 = 5$. The SS2 phase is now completely surrounded by
the SS1* phase and there is no direct transition from the SS2 phase to the
canted-ferromagnetic phase anymore.  The tip of the SS2 lobe can be calculated
analytically from the soft-mode of the SS1* phase (s. appendix
\ref{app:sw_ss1}). It turns out that the value for $J_{\text{tip}}$ has a
remarkably simple form, namely
\begin{equation}
J_{\text{tip}} = U_1\, .
\end{equation}
In the extreme limit of vanishing nearest neighbor interaction, $U_1 =0$, only
the SS1* phase as a supersolid phase survives. The other supersolid phases and
the $3/4$ Mott insulator phase disappear (s. Fig. \ref{pd_0_1}).

{\em (II) Multicritical points and universality classes:} The number of
multicritical points in this parameter range is reduced. For $U_2/U_1 \lesssim
3.1$ there is a tetracritical point at the intersection point of the three
supersolid phases (QC). This multicritical point disappears for $U_2/U_1
\gtrsim 3.1$ and there is a tricritical point at the tip of the SS2 phase
instead. Furthermore there is a multicritical point at the tip of the N\'eel
phase. The properties of all of these multicritical points have already been
discussed in the preceding subsections. There are also no new universality
classes in the phase diagram which have not already been discussed.


\section{Summary and Conclusions}
\label{section:summary}

The analysis of the hard-core Bose-Hubbard model with nearest and next-nearest
neighbor interaction on a square lattice presented in this paper has
demonstrated that the phase diagram can exhibit a rich variety of new quantum
critical and multicritical phenomena. The topology of the phase diagram is
determined by the interplay between the nearest and next-nearest charge
interaction of the bosons and the cubic anisotropy. The principal effect of
these terms is to stabilize several supersolid phases, which are characterized
by the simultaneous presence of superfluidity and charge ordering, where the
latter was described in terms of an Ising and a XY field.

{\em Topology of the phase diagram:} We investigated the whole parameter range
for nearest ($U_1$) and next-nearest neighbor ($U_2$) interaction. By using
ground state energy calculations and stability analysis (spin-wave theory) we
have derived the phase diagrams in mean-field approximation plus small
fluctuations. It turns out that there are three main types of phase diagram
topologies. In parameter regions ($U_1>2U_2$), where the N\'eel ground state is
established at zero field (half-filling), we encounter two supersolid phases,
as already stated by Bruder et al.~\cite{bruder:93}, a superfluid phase, and
three different types of Mott insulating phases. Whereas the spatial order of
the supersolid SS1 phase can be described solely in terms of a checkerboard
pattern, the second supersolid phase (SS2) is a superposition of checkerboard
and stripe domain density waves.  In the region where a stripe ground state at
half-filling is favored ($U_2>U_1/2$) three supersolid phases are found, SS2,
SS3 and SS1*. The SS1* phase is the ``supersolid partner'' of the commensurate
solid with a stripe domain structure and the spatial structure of its boson
density can solely be described in terms of the stripe domain ordering wave
vectors $q_1 = \pi (1,0)$ and $q_2 = \pi (0,1)$. The phase transition from the
SS1* to the SS2 phase is either continuous or first order, depending on the
particular choice of parameters. In terms of an effective action this can be
restated as follows.The phase transition from the SS1* to the SS2 phase is
split up into two phase transitions with an intervening supersolid SS3 phase,
if the Ising field describing the checkerboard structure orders before the
first order transition occurs; otherwise the phase transition is first order.
The transition from the Mott insulator phase $(3/4)$ is of second order into
the SS2. For $U_2<U_1$ the 3/4 phase joints partly (for higher fields) the
canted-ferromagnetic phase; the transition is discontinuous.  The occurrence of
the supersolid phases is strictly coupled to a finite value of $U_2$. Even in
the limit of vanishing nearest neighbor interaction, $U_1=0$, there is the SS1*
supersolid phase present. It is also quite remarkable to note that in the
parameter range $1.149 < U_2/U_1 \lesssim 3$ the transition line from the
canted ferromagnetic to the SS2 phase is the same as for the transition from
the canted ferromagnetic phase to the SS1* phase over the whole range of
hopping matrix elements $J$.

\begin{figure}[hbt]
  \epsfxsize = 0.75\columnwidth \centerline{
    \epsffile{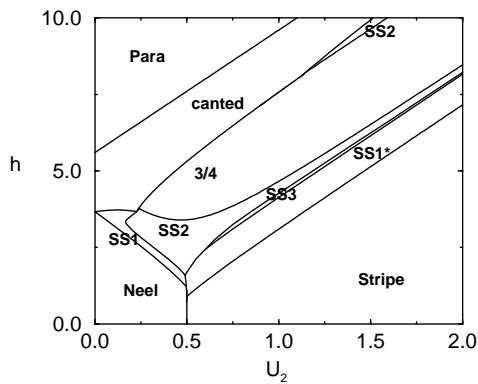} }
\vspace{0.1cm}
\caption{Phase diagram for the hard-core Bose-Hubbard model with $U_1 =1$ and
$J=0.4$ as a function of the next-nearest neighbor interaction $U_2$.}
\label{fig:u2}
\end{figure}

\begin{figure}[hbt]
  \epsfxsize = 0.75\columnwidth \centerline{
    \epsffile{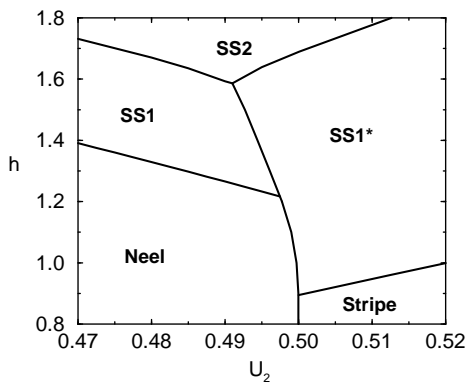} }
\vspace{0.1cm}
\caption{The same as Fig. \ref{fig:u2} but zoomed. The SS1* phase is stable for
slightly smaller values than $U_2 =0.5$, which marks the transition between 
the N\'eel and the stripe phase.}
\label{fig:u2_inset}
\end{figure}

Figs.~\ref{fig:u2}-\ref{fig:u2_inset} show the complex topology of the phase
diagram for fixed hopping matrix elements but varying strength of the
next-nearest neighbor interaction.

In investigating the critical and multicritical behavior in the various
parameter regions of the hard-core Bose-Hubbard model we have also found a rich
variety of critical and multicritical phenomena.  

{\em Superfluid to supersolid transition:} There are three different
universality classes for the phase transition from the superfluid to the
supersolid phases. These are the XSS-SF, CSS-SF and MSS-SF universality class
for the transition from the superfluid (SF) to the checkerboard supersolid
(SS1), stripe domain supersolid (SS1*) and supersolid with an intermediate
boson density profile (SS2). The CSS-SF transition is in the universality
classes of the $D=2+1=3$ dimensional XY model. The critical fluctuations
do not lead to deviations from Bose-liquid behavior at the transition. The
critical phenomena at the XSS-SF transition are more interesting since it can
either be fluctuation driven first order or be governed by a strong coupling
fixed point implying that the superfluid component displays ``non Bose-liquid''
behavior.  For phase diagrams of type I topology the phase transition from the
superfluid to the mixed supersolid phase is always first order. Starting with
$U_2 \approx U_1/2$ (type II topology) one finds that within mean-field theory
part of this phase boundary becomes second order and a tricritical point
appears. The critical behavior at the second order line constitutes a new
MSS-SF universality class, which is interesting in several respects. First of
all, the trilinear coupling between the checkerboard and stripe domain order
parameter induces criticality even for a massive checkerboard field at the
phase boundary where the mass of the stripe domain field becomes zero. Second,
the critical exponent for the ``slaved'' checkerboard field $\phi$ are
anomalously large already at the mean-field level. For instance one finds
$\phi_{\text{mf}} \propto \langle \psi \rangle ^2$, which implies
$\beta_\phi^{\text{mf}} = 1$. We suppose that a detailed analysis
\cite{frey-pich:97} of this new universality class will reveal interesting
critical anomalies; in particular there will be cusp-like singularities in the
``slaved'' checkerboard field.

{\em Supersolid to supersolid transition:} The phase transitions between the
various supersolid phases can either be first or second order. If they are
second order, they all belong to the XSS-SF universality class. In the
parameter range where the checkerboard solid is the ground state at
half-filling (type I topology) there is a first order transition from the
checkerboard supersolid (SS1) to the mixed supersolid phase (SS2) at large
fields (fillings). Lowering the field the transition becomes continuous at the
tip of the SS2 phase. The critical behavior at this point is described by a new
quantum tricritical model different from the $D=3$ classical Ising model. In
the parameter range where the stripe domain solid is the ground state at
half-filling there is no direct second order transition from the stripe domain
supersolid (SS1*) to the mixed supersolid (SS2); it is always first order.  At
certain regions of parameter space, however, there is an intermediate
supersolid phase (SS3) and the first order line splits into two second order
phase boundaries (both are in the XSS-SF universality class). At the tip of the
intermediate SS3 phase there is a new bicritical quantum model. The effective
action contains two Ising fields which are coupled by an anisotropic trilinear
vertex. The structure of the effective action shows some resemblance with
anisotropic Potts models.

{\em Supersolid to commensurate solid transition:} The phase transitions from
the supersolid to the neighboring commensurate solid phases are all most likely
mean-field like with logarithmic corrections in $D=2+1$ dimensions.

{\em Multicritical points:} For finite values of the hopping matrix element we
have found up to seven different multicritical points on a single phase
diagram. Some of them are novel type of quantum multicritical points. In
particular, we found a new quantum tricritical point at the tip of the mixed
supersolid phase. In the vicinity of these tricritical points there are
critical end points with critical properties which are an interesting field for
future investigations. Especially the critical end point with a critical
``spectator phase'' might engender novel ``non Bose-liquid'' behavior in the
spectating phase.

{\bf Acknowledgment:} It is a pleasure to acknowledge helpful discussions with
Leon Balents, Markus Hummel, Uwe T\"auber and Peter Young. This work was
supported by the Deutsche Forschungsgemeinschaft (DFG) under Contract 
No.\ FR\ 850/2 and No.\ PI\ 337/1-1.

\appendix

\section{Spin-wave analysis of the 3/4 phase}
\label{app:sw_3/4}

In this section we present the main results of the spin-wave analysis of the
$3/4$ phase. Due to the four-sublattice structure we introduce four different
Boson creation and annihilation operators $(a_l^\dagger, a_l), (b_l^\dagger,
b_l), (c_l^\dagger, c_l), (d_l^\dagger, d_l)$. By using the linearized
Holstein-Primakoff transformation for the first three with spin up ($l\in {\cal
  L}_1, {\cal L}_2$ and ${\cal L}_3$) and the fourth spin down ($m\in {\cal
  L}_4$),
\begin{mathletters} 
\begin{eqnarray}
S_l^z & = & S-a_l^\dagger a_l \\
S_l^x & = & \sqrt{S\over 2}\left( a_l+a_l^\dagger\right)\\
S_l^y & = & -i\sqrt{S\over 2}\left( a_l-a_l^\dagger\right)\\
S_m^z & = & -S+d_l^\dagger d_l \\
S_m^x & = & \sqrt{S\over 2}\left( d_l+d_l^\dagger\right)\\
S_m^y & = & -i\sqrt{S\over 2}\left( d_l-d_l^\dagger\right)\, ,
\end{eqnarray}
\end{mathletters}
and inserting in the Hamiltonian (Eq.~\ref{XXZ}) we derive the following
Hamiltonian:
\begin{eqnarray}
{\cal H} &=& -{NSh\over 2} \nonumber \\
&&+ \sum_q \Bigl[ A^{(1)}a_q^\dagger a_q +A^{(2)}b_q^\dagger
b_q \nonumber \\
&&~+ A^{(1)}c_q^\dagger c_q + A^{(4)}d_q^\dagger d_q \nonumber\\
&&~+B_q(a_qb_q^\dagger +a_q^\dagger b_q+c_qd_{-q}+c_q^\dagger
d_{-q}^\dagger)\nonumber\\
&&~+C_q(a_qd_{-q}+a_q^\dagger
d_{-q}^\dagger +b_qc_q^\dagger +b_q^\dagger c_q) \Bigr] 
\end{eqnarray}
with the coefficients
\begin{mathletters} 
\begin{eqnarray}
A^{(1)} & = & h-8SU_2 \\
A^{(2)} & = & h+8S(U_2-U_1) \\
A^{(4)} & = & -h+8S(U_1+U_2) \\
B_q & = & -4SJ \cos{q_x} \\
C_q & = & -4SJ \cos{q_y}\, .
\end{eqnarray}
\end{mathletters}
This bilinear Hamiltonian can be diagonalized by standard method, e.g. by
introducing proper Greens functions. The excitation spectrum is then given by
the zero of the following determinant
\begin{equation}
\left(
\begin{array}{cccc}
A^{(1)}-\omega & B_q & C_q & 0\\
B_q & A^{(2)}-\omega & 0 & C_q\\
C_q & 0 & A^{(4)}+\omega & B_q\\
0 & C_q & B_q & A^{(3)}-\omega
\end{array}
\right)
\end{equation}
As a result, we get a polynomial of fourth order in the excitation energy
\begin{equation}
\omega^4 +r\omega^3+s\omega^2+t\omega +u = 0
\end{equation}
with coefficients
\begin{mathletters}
\begin{eqnarray}
r & = & A^{(4)}-A^{(2)}-2A^{(1)}\\
s & = & {A^{(1)}}^2+2A^{(1)}A^{(2)}-2A^{(1)}A^{(4)}-A^{(2)}A^{(4)}\\
t & = & 2A^{(1)}A^{(2)}A^{(4)}+{A^{(1)}}^2(A^{(4)}-A^{(2)})\nonumber\\
&&-(B_q^2-C_q^2)(A^{(2)}+A^{(4)})\\
u & = &
-{A^{(1)}}^2A^{(2)}A^{(4)}+A^{(1)}(A^{(2)}+A^{(4)})(B_q^2+C_q^2)\nonumber\\
&& -(B_q^2-C_q^2)^2\, .
\end{eqnarray}
\end{mathletters}
This polynomial defines the four branches of the excitation spectrum.
Calculating the soft-mode at $q=0$ leads to the phase boundaries of the $3/4$
phase. Due to $C_0=B_0$ the spectrum can be simplified: the first solution is
$\omega = A^{(1)}$, independent of $q$. The other solutions are given by a
third order polynomial given by Eq.~\ref{sw_3/4}.  For $J=0$ the solution are
given by $A^{(1)},A^{(2)}$ and $A^{(4)}$ from which the boundaries can be
readily seen.

\section{Spin-wave analysis of the SS1 and SS1* phase}
\label{app:sw_ss1}

These phases have two-sublattice spin structure which can be characterized by
the wave vectors $\tilde q$ which is $q_0$ for the SS1 phase and $q_1$ and
$q_2$ for the SS1* phase. Spin-wave calculation is done via the linearized
transformation in Fourier space (Eq.~\ref{Trafo}). The biquadratic term in the
Bose operators can be written in the form \cite{pich94:PhD}
\begin{eqnarray}
\hat H^{(2)}
&=& \sum_{q}
    \Bigl[ A_{q}~a_{q}^{\dagger}a_{q}
         + \frac{B_{q}}{2} \left( a_{q}~a_{-q} + 
                                a_{q}^{\dagger}a_{-q}^{\dagger}
                         \right) \nonumber \\
&&~~~~+ C_{q} \left( a_{q}~a_{-{q}-\tilde {q}} +
                     a_{q}^{\dagger}a_{-{q}-\tilde {q}}^{\dagger}
              \right) \nonumber \\
&&~~~~+ D_{q} \left( a_{q}^{\dagger}a_{{q}+\tilde {q}} +
                     a_{{q}+\tilde {q}}^{\dagger}a_{q}
              \right) \Bigr]
\end{eqnarray}
with coefficients
\begin{mathletters}
\begin{eqnarray}
A_{q} 
&=& S \Bigl[ 2J_0-U_{q}-J_{q} \nonumber\\
&&+\bigl( 2J_{\tilde {q}}+2J_0-2U_{\tilde {q}}-2U_0 \nonumber \\
&&-U_{q+\tilde {q}}-U_{q}+ J_{q+\tilde {q}}+J_{q}
    \bigr)  \sin^2\gamma\sin^2\delta \nonumber\\
&&+\bigl( 2U_{\tilde {q}} -J_{{q}+\tilde {q}} -2J_0 +U_{q}
   \bigr) \sin^2\delta \nonumber\\
&&+\bigl( 2U_0-2J_0-J_{q}+U_{q} \bigr) \sin^2\gamma  
\Bigr] \nonumber \\
&&+ h\sin\gamma\cos\delta \\
B_{q} &=& S \Bigl[ U_{q}-J_{q} \nonumber \\
&&+\bigl( U_{q}+U_{{q}+\tilde {q}}-J_{q}
         -J_{{q}+{q}_0} 
   \bigr)  \sin^2\gamma\sin^2\delta \nonumber\\
&&+\bigl( J_{{q}+\tilde {q}}-U_{q} 
   \bigr)  \sin^2\delta
  +\bigl( J_{q}-U_{q} \bigr)  \sin^2\gamma\Bigr] \nonumber \\
C_{q} &=& S \bigl( J_{q}-U_{q} \bigr) 
              \cos\gamma\sin\delta\sin\gamma\cos\delta \\
D_{q} &=& S \Bigl[
          \bigl( U_{\tilde {q}}+U_0 -J_{\tilde {q}}
                -J_0+U_{q}-J_{q}
          \bigr) \nonumber \\
&&~~~\times\cos\gamma\sin\delta\sin\gamma\cos\delta \Bigr] \nonumber\\
&&+ \frac{h}{2} \cos\gamma\sin\delta .
\end{eqnarray}
\end{mathletters}
Here the angles $\delta$ and $\gamma$ are the difference and sum angles of the
spins in respect to the magnetic field (Eq.~\ref{equi-ss1} and
Eq.~\ref{equi-ss1*}). The excitation spectrum for this Hamiltonian then yields
\begin{equation}
E_{q}^{(i)} = \sqrt{(\Omega_1 \pm \Omega_2)/2},
\label{Magnon}
\end{equation}
with
\begin{mathletters}
\begin{eqnarray*}
\Omega_1 
&=& A_{q}^2-B_{q}^2+A_{{q}+\tilde {q}}^2
   -B_{{q}+\tilde {q}}^2 
    \nonumber \\
&&-2(C_{q}+C_{{q}+\tilde {q}})^2
  +2(D_{q}+D_{{q}+\tilde {q}})^2, \\
\Omega_2^2 
&=& \left( A_{q}^2 - B_{q}^2 - A_{{q}+\tilde {q}}^2
         + B_{{q}+\tilde {q}}^2 \right)^2 \nonumber \\
&&+ 4 \left( D_{q} + D_{{q}+\tilde {q}}^2 \right) 
            \left[ \left( A_{q}+A_{{q}+\tilde {q}} \right)^2
                   -B_{q}^2-B_{{q}+\tilde {q}}^2 \right] 
  \nonumber \\
&&- 4 \left( C_{q} + C_{{q}+\tilde {q}} \right)^2
            \left[ \left( A_{q}-A_{{q}+\tilde {q}} \right)^2
                   -B_{q}^2-B_{{q}+\tilde {q}}^2 \right]
  \nonumber \\
&&+ 8 \Bigl\{ B_{{q}+\tilde {q}} B_{q}
             \left[ \left(D_{q}+D_{{q}+\tilde {q}} \right)^2
                   +\left(C_{q}+C_{{q}+\tilde {q}}\right)^2 
             \right]
  \nonumber \\
&&- 2 \left( C_{q}+C_{{q}+\tilde {q}} \right)
      \left( D_{q}+D_{{q}+\tilde {q}} \right)
      \nonumber \\
&&~\times\left( A_{q}B_{{q}+\tilde {q}}
            +A_{{q}+\tilde {q}}B_{q}
      \right) \Bigr\}.
\end{eqnarray*}
\end{mathletters}
For the SS1 phase, $\tilde q=q_0$, a soft-mode appears at $q_1$ and $q_2$. 
The spectrum for this wave vector can be simplified and vanishing of the second
branch, $E_q^{(2)}$, leads to the condition
\begin{equation}
A_{q_1}-B_{q_1}-2(D_{q_1}-C_{q_1}) = 0\, ,
\end{equation}
which defines with help of the angles $\gamma$ and $\delta$ a cubic equation in
the field $h$. The tip of the SS2 lobe then is given by that point where two
solutions are degenerated, i.e. where the discriminant vanishes.

In the striped phase SS1*, $\tilde q = q_1,q_2$, there is a soft-mode at the
checkerboard wave vector $q=q_0$. In this case the spectrum becomes a quartic
function in the field and the tip of the SS2 lobe can be found by
studying the corresponding cubic resolvent.

\end{multicols} 


\begin{thebibliography}{10}
\bibitem{note} Note, however, that the stability of the free energy requires
  constraints on the parameters $u$ and $v$. Furthermore, in examining the flow
  diagram of the cubic model within $\varepsilon$-expansion one finds that
  there are two regions in the $(u,v)$-plane which lie outside the domain of
  attraction of the stable fixed point (for a full discussion see e.g.  Ref.
  \protect{\cite{nattermann-trimper:75,lyuksyutov-prokrovskii:75,lyuksyutov:76}}).
  Adding a quadratic symmetry breaking term, $g \psi_1 \psi_2$, even the first
  order transition (associated with the parameter regions of the $(u,v)$-plane
  which lie outside the domain of attraction of the stable fixed point) can
  become continuous. The phase diagram turns out to be rather complex with
  various interesting multicritical
  points~\protect{\cite{kerszberg-mukamel:81}}.
\bibitem{note_tetra}
If one introduces new fields $\sigma = \psi_1 + \psi_2$ and $\pi = \psi_1 -
  \psi_2$ one obtains an effective action with a uniaxial anisotropy (induced
  by the quadratic symmetry breaking field $g$) and a cubic symmetry breaking
  field $\tilde v$. Then the effective action discussed here closely resembles
  the structure of a free energy discussed Bruce and Aharony
  \protect\cite{bruce-aharony:75} in the context of classical critical
  phenomena. They find that for number of spin components $n<n_c (D)$, where
  $n_c(D)$ is the critical number of components above which the cubic fixed
  point becomes stable, the critical lines separating the intermediate phase
  from the other two phases merge into a single ``flop'' line making the
  tetracritical point apparently bicritical. Similar conclusion should hold
  here but differences could result due to the coupling to the superfluid
  phase.
\end{thebibliography}
\end{document}